\documentclass[preprint]{aastex}
\usepackage{emulateapj5}
\usepackage{apjfonts}
\usepackage{natbib}
\usepackage{epsfig}
\input psfig.tex

\def\aas{{ A\&AS}}
\def\aj{{AJ}}

\def\annrev{{ARA\&A}}
\def\apj{{ApJ}}
\def\apjs{{ApJS}}
\def\asec{$^{\prime\prime}$}

\def\hst{{\it HST}}

\def\lax{{$\mathrel{\hbox{\rlap{\hbox{\lower4pt\hbox{$\sim$}}}\hbox{$<$}}}$}}
\def\gax{{$\mathrel{\hbox{\rlap{\hbox{\lower4pt\hbox{$\sim$}}}\hbox{$>$}}}$}}
\def\simlt{\lower.5ex\hbox{$\; \buildrel < \over \sim \;$}}
\def\simgt{\lower.5ex\hbox{$\; \buildrel > \over \sim \;$}}

\def\mbh{{$M_{\rm BH}$}}

\def\mnras{{MNRAS}}

\def\percm2{cm$^{-2}$}

\def\lbul{$L_{\rm bul}$}
\def\mbul{$M_{\rm bul}$}
\def\ser{S\'{e}rsic}
\def\hnr{$L_{\rm host}/L_{\rm nuc}$}
\def\vel{$\sigma_{*}$}
\def\nsn{$(S/N)_{\rm nuc}$}

\slugcomment{To Appear in {\it
The Astrophysical Journal Supplement Series}.}
\shorttitle{Decomposition of AGN Host Galaxies}
\shortauthors{KIM et al.}

\begin{document}

\title{Decomposition of the Host Galaxies of Active Galactic Nuclei Using 
{\it Hubble Space Telescope}\ Images}
\author{Minjin Kim\altaffilmark{1,2}, Luis C. Ho\altaffilmark{1}, Chien Y.
Peng\altaffilmark{3}, Aaron J. Barth\altaffilmark{4}, and
Myungshin Im\altaffilmark{2}}

\altaffiltext{1}{The Observatories of the Carnegie Institution of Washington,
813 Santa Barbara Street, Pasadena, CA 91101; mjkim@ociw.edu, lho@ociw.edu.}

\altaffiltext{2}{Department of Physics and Astronomy, Frontier Physics
Research Division (FPRD), Seoul National
University, Seoul, Korea; mim@astro.snu.ac.kr.}

\altaffiltext{3}{NRC Herzberg Institute of Astrophysics, 5071
West Saanich Road, Victoria, British Columbia, Canada V9E2E7;
cyp@nrc-cnrc.gc.ca.}

\altaffiltext{4}{Department of Physics and Astronomy, University of
California at Irvine, 4129 Frederick Reines Hall, Irvine, CA 92697-4575;
barth@uci.edu.}

\begin{abstract}
Investigating the link between supermassive black hole and galaxy evolution
requires careful measurements of the properties of the host galaxies. We
perform simulations to test the reliability of a two-dimensional 
image-fitting technique to decompose the host galaxy and the active galactic 
nucleus (AGN), especially on images obtained using cameras onboard the 
{\it Hubble Space Telescope (HST)}, such as the Wide-Field Planetary Camera 2, 
the Advanced Camera for Surveys, and the Near-Infrared Camera and Multi-Object 
Spectrometer.  We quantify the relative importance of spatial, temporal, and 
color variations of the point-spread function (PSF).  To estimate uncertainties 
in AGN-to-host decompositions, we perform extensive simulations that span a 
wide range in AGN-to-host galaxy luminosity contrast, signal-to-noise ratio, 
and host galaxy properties (size, luminosity, central concentration).  We find 
that realistic PSF mismatches that typically afflict actual observations
systematically lead to an overestimate of the flux of the host galaxy.  Part 
of the problem is caused by the fact that the {\it HST}\ PSFs are 
undersampled.  We demonstrate that this problem can be mitigated by broadening 
both the science and the PSF images to critical sampling without loss of 
information.  Other practical suggestions are given for optimal analysis of 
\hst\ images of AGN host galaxies.
\end{abstract}


\section{Introduction}

The mass of supermassive black holes (BHs) is strongly correlated with the
luminosity (\lbul; Kormendy \& Richstone 1995; Magorrian et al. 1998) and the 
stellar velocity dispersion (\vel; Gebhardt et al. 2000; Ferrarese \& Merritt 
2000) of the bulge of the host galaxy.  These scaling relations are often 
interpreted to be evidence that central BHs and their host galaxies are 
closely connected in their evolution (see reviews in Ho 2004).  The empirical 
correlations between BH mass and host galaxy properties can even be used 
as tools to track the progress of mass assembly during galaxy evolution 
(Peng et al. 2006a, 2006b; Woo et al. 2006; Ho 2007b).  
The central BH mass correlates most strongly with the bulge component of a 
galaxy rather than with its total mass or luminosity.  This has motivated 
detailed bulge-to-disk decompositions of the host galaxies to better quantify 
the intrinsic scatter of the \mbh-\mbul\ and \mbh-\lbul\ relations in the local 
Universe (Marconi \& Hunt 2003; H\"{a}ring \& Rix 2004).  

To probe when the BH-host galaxy relations were established and how they 
evolved, it is of paramount importance to extend similar studies out to higher 
redshifts.  However, direct measurement of BH mass based on spatially resolved 
stellar or gas kinematics is unfeasible for all but the nearest galaxies with 
low levels of nuclear activity.  Accessing BHs and their host galaxies at 
cosmological distances requires a different approach---one that relies on 
active galactic nuclei (AGNs).  The masses of BHs in type~1 (unobscured, 
broad-line) AGNs can be readily estimated with reasonable accuracy using the 
virial technique with single-epoch optical or ultraviolet spectra (Ho 1999; 
Wandel et al. 1999; Kaspi et al.  2000; Greene \& Ho 2005b; Peterson 2007).  
Quantitative measurements of the host galaxies with active nuclei, on the 
other hand, are less straightforward to obtain because the presence of the AGN 
introduces significant practical difficulties, as well as potential biases.  
This is especially problematic for the bulge component of the host, which is 
maximally affected by the bright AGN core.  A variety of techniques have 
been employed, both kinematical (e.g., Nelson et al. 2004; Onken et al. 2004; 
Barth et al. 2005; Greene \& Ho 2005a, 2006; Ho 2007; Salviander et al. 2007; 
Ho et al. 2008; Shen et al. 2008) and photometric (e.g., McLure \& Dunlop 
2002; Peng et al. 2006b; Greene et al. 2008).
\begin{figure*}
\psfig{file=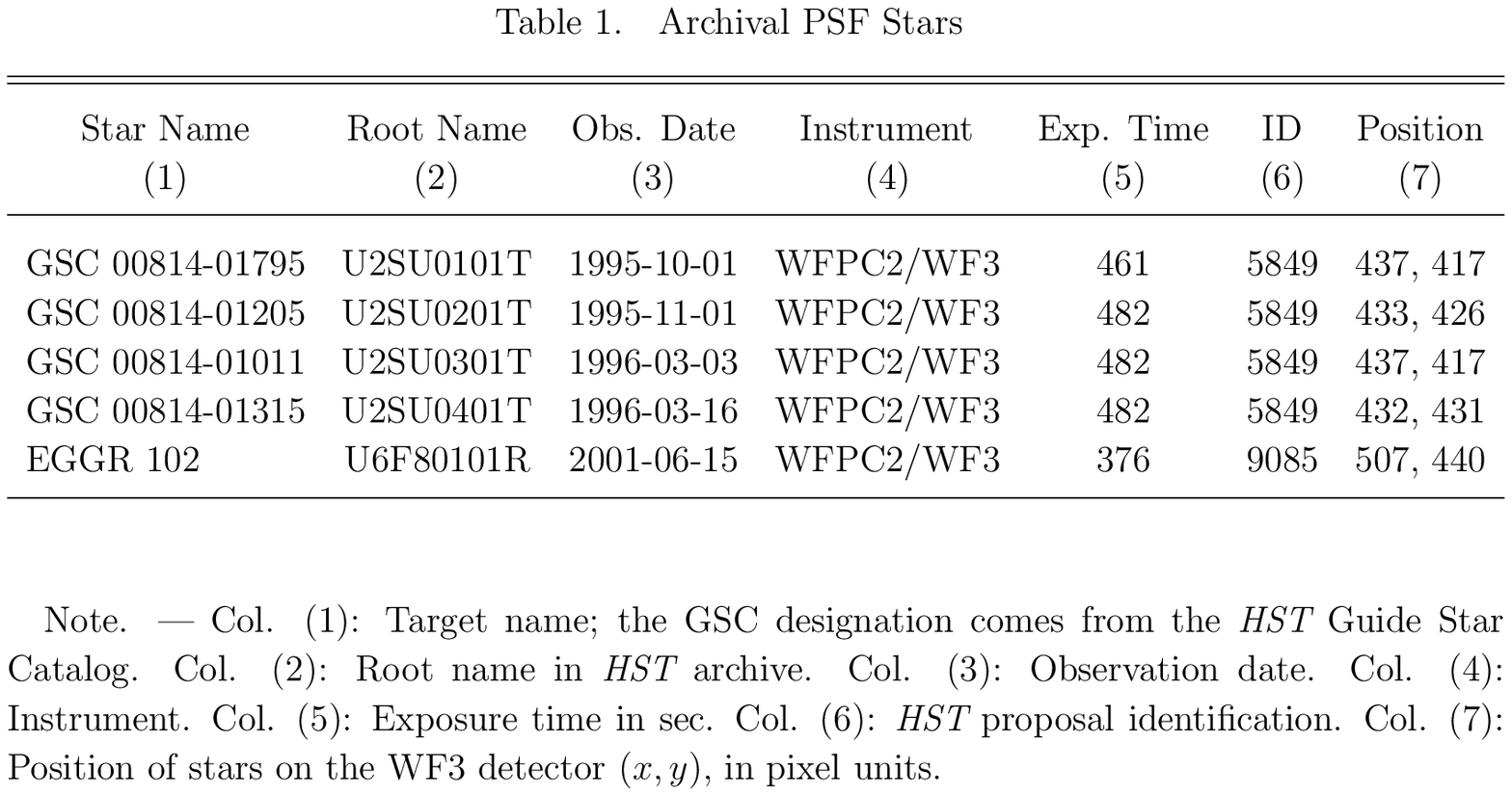,width=18.5cm,angle=0}
\vskip -0.4cm
\end{figure*}

The main challenge for photometric studies of active galaxies lies in 
separating the central AGN light from the host galaxy. While 
structural decomposition can be done by fitting analytic functions to the 
one-dimensional (1-D) light distribution, two-dimensional (2-D) analysis makes 
maximal use of all the spatial information available in galaxy images 
(Griffiths et al. 1994; Byun \& Freeman 1995; de~Jong 1996; Wadadekar et al. 
1999; Peng et al. 2002; Simard et al. 2002; de~Souza 
et al. 2004) and thus provides the most general and most robust method to 
decouple image subcomponents.  This flexibility proves to be especially 
important in the case of active galaxies, where the contrast between the 
central dominant point source and the underlying host can be very high.  In 
this regime, achieving a reliable decomposition requires knowing the 
point-spread function (PSF) to high accuracy.  

This paper discusses the complications and systematic effects involved in 
photometric decomposition of AGN host galaxies, especially as it applies to 
images taken with the {\it Hubble Space Telescope (HST)}.  Although numerous 
{\it HST} studies of AGN hosts have been published, very few have explicitly 
investigated the systematic uncertainties or practical limitations of host 
galaxy decomposition.  We make use of an updated version of the 2-D 
image-fitting code GALFIT (Peng et al. 2002) to generate an extensive 
set of simulated images of active
galaxies that realistically mimic actual 
{\it HST} observations.  We then apply the code to fit the artificial images 
and quantify how well we can recover the input parameters for the AGN and for 
the host galaxy.  Our strategy resembles those employed in the quasar host 
galaxies studies by Jahnke et al. (2004) and S\'{a}nchez et al. (2004), except 
that we have optimized our simulations to be applicable to nearby bright AGNs, 
a regime of most interest to us (Kim et al. 2008).  
Nearby bright AGN hosts observed using coarse pixels and low exposure time can
be equally challenging to analyze as high-z AGNs observed with high resolution
and signal-to-noise.  In other words, fundamentally, the difficulty of the
analysis depends only on the following 3 relative conditions: angular
resolution of the detector vs. the object being studied (i.e. object scale
size in pixels), AGN-to-host contrast, and overall object signal-to-noise.
Since we cover a wide range of parameter space the simulations described here
should be applicable to most {\it HST}\ imaging studies of AGN hosts,
including distant quasars.
We hope that the results of this study can serve as a guide to other 
investigators working with similar data.

We describe the details of PSF variations in \S{2}. In \S{3}, we present the
simulation procedure and our results from fitting artificial images using
PSF models with varying degrees of realism.  A discussion and 
summary of the main results are given in \S{4}.

\section{GENERAL PSF CONSIDERATIONS}

Our primary simulations are designed to test observations using the Wide Field 
(WF) camera of the the Wide-Field and Planetary Camera 2 (WFPC2) onboard \hst.
The largest studies to date of AGN hosts have been done with the WF (e.g., Bahcall 
et al. 1997; McLure et al. 1999; Floyd et al. 2004).  While we concentrate on 
understanding observations made with the WF, we also consider the other main 
imaging instruments onboard \hst\ (\S{3.7}).  As we will see, the behavior for 
the other cameras is both qualitatively and quantitatively similar.

First and foremost, accurate decomposition of AGN host galaxies requires the 
analysis PSF to be stable and well-matched to the science image.  The 
stability of the PSF depends on
a number of instrumental and environmental factors.  For instance, the PSF
shape may depend sensitively on the telescope focus, detector temperature,
optical distortion, relative alignments of the telescope optics, and telescope
jitter.  Therefore, changes in atmospheric conditions or focus
drifts may cause the PSF to vary with time.  Optical distortions in the 
telescope optics can also cause the PSF shape to change subtly 
across the field of view of the detector.  
Ground-based observations are most susceptible to these effects.

In most respects, the \hst, in the near absence of terrestrial environmental 
effects, produces the most stable PSF of all current optical/near-infrared 
telescopes.  It thus has been an incomparable choice for studying AGN host 
galaxies (e.g., Bahcall et al.  1997; Boyce et al.  1998; McLure et al.  1999; 
Schade et al.  2000; Dunlop et al.  2003).  Nevertheless, subtle changes in 
the telescope structure and instruments, such as plate-scale ``breathing'' or 
focus changes due to out-gassing of the telescope structure, do cause the 
PSF to vary slowly with time (e.g., Jahnke et al.  2004; S\'{a}nchez et al. 
2004; Kim et al. 2007).  The shape of the PSF also changes because the PSF 
is undersampled (this issue is discussed in detail in \S{2.2}). 

PSF variability, which produces a mismatch between the science data and the
analysis PSF, is the leading cause for systematic measurement errors in AGN 
image analysis.  When PSFs are fitted to AGNs,
mismatches produce residuals in 
the image that do not obey Poisson statistics.  If the AGN is sufficiently
bright, the systematic residuals exceed the random noise and may rival the
light from the host galaxy beneath.

To study the effects of PSF mismatch on real data, we will first identify the
most important causes for PSF mismatches and then quantify their relative
importance.  We use two types of PSF images.  The first are synthetic PSFs 
from TinyTim\footnote{http://www.stsci.edu/software/tinytim/tinytim.html} (Krist 1995).  We use this 
program to create several non-oversampled PSF images, each of size $50 \times 
50$ pixels, an area large enough to include $>$99\% of the total flux. The PSF 
is exactly centered on a pixel, and we do not introduce jitter into the model, 
which has the same practical effect as PSF mismatch that we discuss later. 
In addition, we also use PSF images derived from actual observations of five
stars contained in the \hst\ archive (Table 1).
\begin{figure*}
\psfig{file=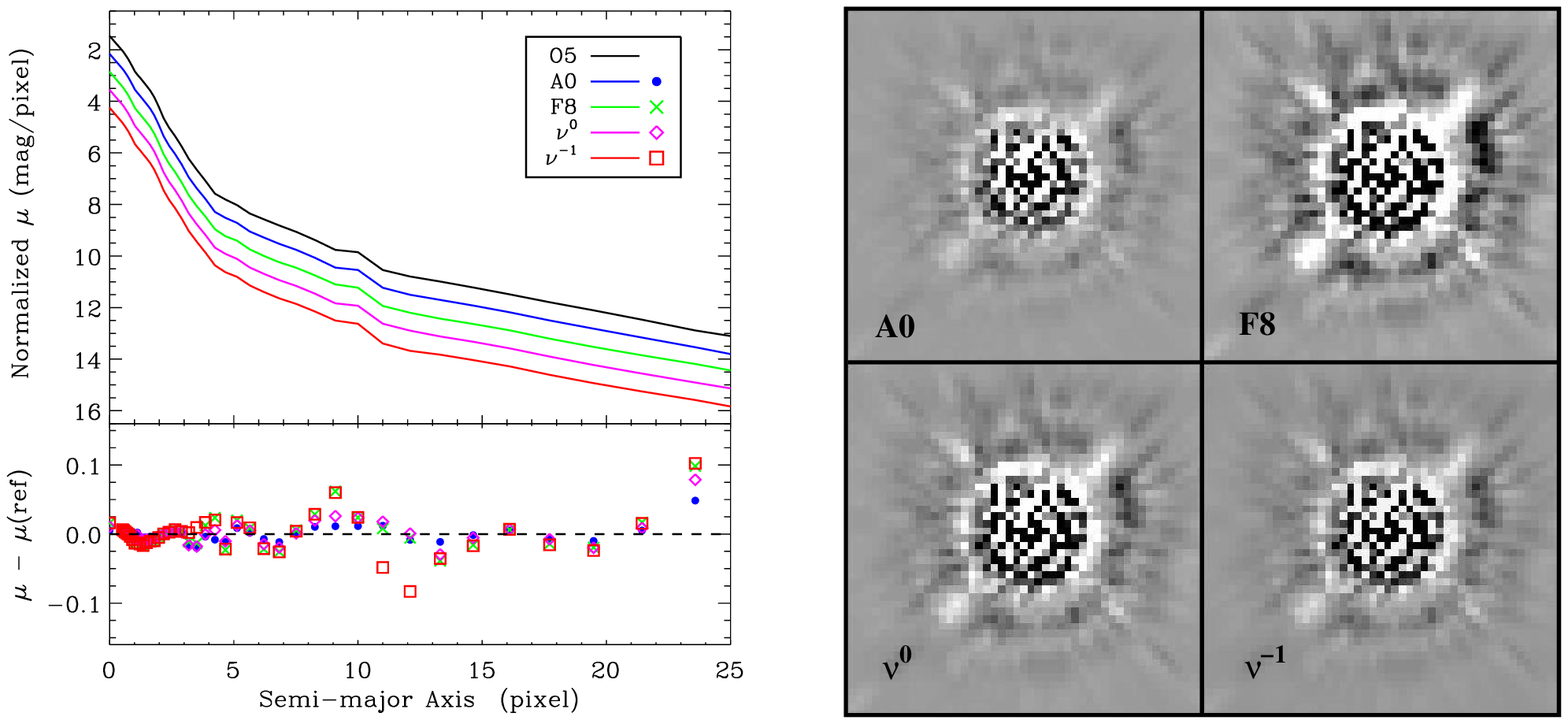,width=18.5cm,angle=0}
\figcaption[fig1.ps]
{
PSF variation as a function of spectral type.  ({\it Left})
PSFs in the F606W filter created by TinyTim with different SEDs: O5 star ({\it
black line}, the reference PSF), A0 star ({\it blue line and circles}),
F8 star ({\it green line and crosses}), $F_{\nu} \propto \nu^{0}$ power law
({\it magenta line and diamonds}), and $F_{\nu} \propto \nu^{-1}$ power law
({\it red line and squares}).  The top panel shows the normalized surface
brightness profiles (each offset by 0.7 units in the ordinate for clarity) as
a function of semi-major axis.  The bottom panel shows the residuals of the
surface brightness profiles between the reference PSF and the other PSFs.
({\it Right}) Residual images of each PSF image after fitting it to the
reference PSF.  The size of each box is 50$\times$50 pixels
($\sim 5$\asec$\times$5\asec).  All images are on a linear gray scale.
The variation is almost negligible.
\label{fig1}}
\vskip 0.3cm
\end{figure*}
%

\subsection{PSF Variation}

PSF variations in \hst\ images can be broadly separated into three categories:
color differences due to the spectral energy distribution (SED) of an 
astronomical object, spatial changes due to optical distortions across the 
field of view, and temporal changes due to gradual movements of the telescope 
optical system.

 We first compare these three effects to see which of them 
produces the largest PSF changes.

{\it Color variability} \ \ \ \ \ 
The PSF used in the analysis (e.g., a star) often has a different color than 
the science target (e.g., an AGN).  The color differences may translate into 
PSF mismatches because \hst\ produces diffraction-limited images.  To study 
this effect, we can use TinyTim PSFs because chromatic differences of light 
propagation through the telescope optics should be fairly accurately known.  
We create several non-oversampled PSFs with different SEDs in the F606W filter,
which is the widest filter and will have the strongest chromatic PSF variation. 

Figure 1 summarizes our results in terms of the average 1-D surface brightness 
profiles of the models, extracted using the IRAF\footnote{IRAF (Image 
Reduction and Analysis Facility) is distributed by the National Optical 
Astronomy Observatories, which are operated by AURA, Inc., under cooperative 
agreement with the National Science Foundation.} task {\it ellipse}.  The 
variation due to SED is less than 10\% at 
all radii, which is negligible compared to that due to position difference 
(see below).  We also perform the same test with different filters 
(e.g., F555W and F814W) and find the variation to be still minimal.  These 
results should be robust to the extent that the filter traces and detector 
responses are well known.  We caution, however, that the differences may be 
more prominent if the filters suffer from red or blue light leaks, such as 
that seen in the ACS/HRC F850LP filter (Sirianni et al. 2005).  Furthermore, 
we do not suggest that the color effect should be ignored when it is possible 
to mitigate it in the observations.  Its relative importance, however, is small
compared to spatial and temporal variability. In general, white dwarf standards
like EGGR102 are a reasonable color match to quasar nuclei. 

{\it Spatial variability} \ \ \ \ \ We create several PSFs, located at
different positions on the WF3 detector on WFPC2, again using TinyTim.  The one
located at the center of the chip, at pixel position ($x,y$) = (400,400), is
used as the point of reference.  The others are generated at other image 
positions, with otherwise the same conditions as the reference PSF.  
It is known that the 
TinyTim software may not perfectly model real stellar images, in absolute 
terms, due to light scattering, spacecraft jitter, focus, and other effects.  
However, external factors largely affect the PSF across the detector 
uniformly, whereas it is the differential optical distortions due to geometric 
optics (center vs. outskirts of the field) that concern us here.  In 
general, global field distortions due to geometric optics are both very well 
known and stable.  Even if otherwise, we need not have high accuracy for our 
purposes.  Therefore, the TinyTim synthetic PSFs more than suffice for our 
goals.
\begin{figure*}
\psfig{file=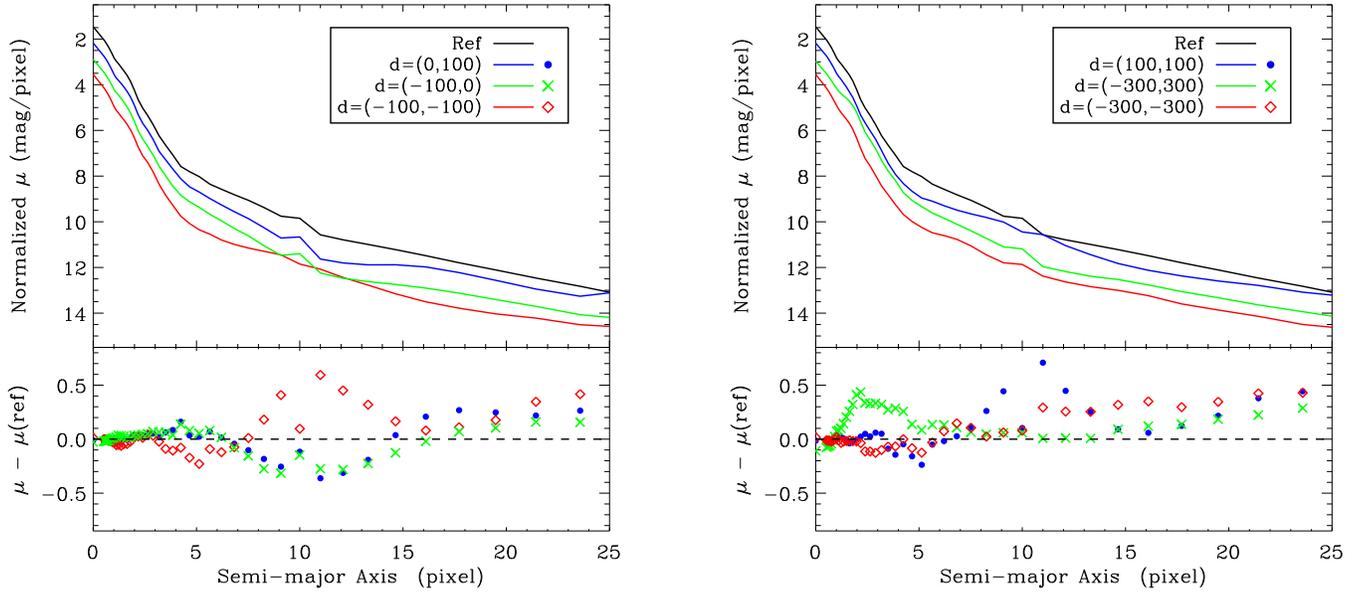,width=18.5cm,angle=0}
\figcaption[fig2.ps]
{
Similar to left panel of Fig. 1, except here we show PSF variations due to
position changes.  The reference PSF ({\it black line}) is located at the
center, ($x,y$) = (400,400) of the WF3 detector on WFPC2. The other PSFs are
created at different locations, with the following pixel offsets with respect
the reference PSF.  ({\it Left}) ($x,y$) = (0,100) ({\it blue line and
circles}), ($-$100,0) ({\it green line and crosses}), and ($-$100,$-$100)
({\it red line and diamonds}).  ({\it Right}) ($x,y$) = (100,100) ({\it blue
line and circles}), ($-$300,300) ({\it green line and crosses}), and
($-$300,$-$300) ({\it red line and diamonds}).  As the position offset
increases, the discrepancy between the PSFs increases.
\label{fig2}}
\end{figure*}
\begin{figure*}
\psfig{file=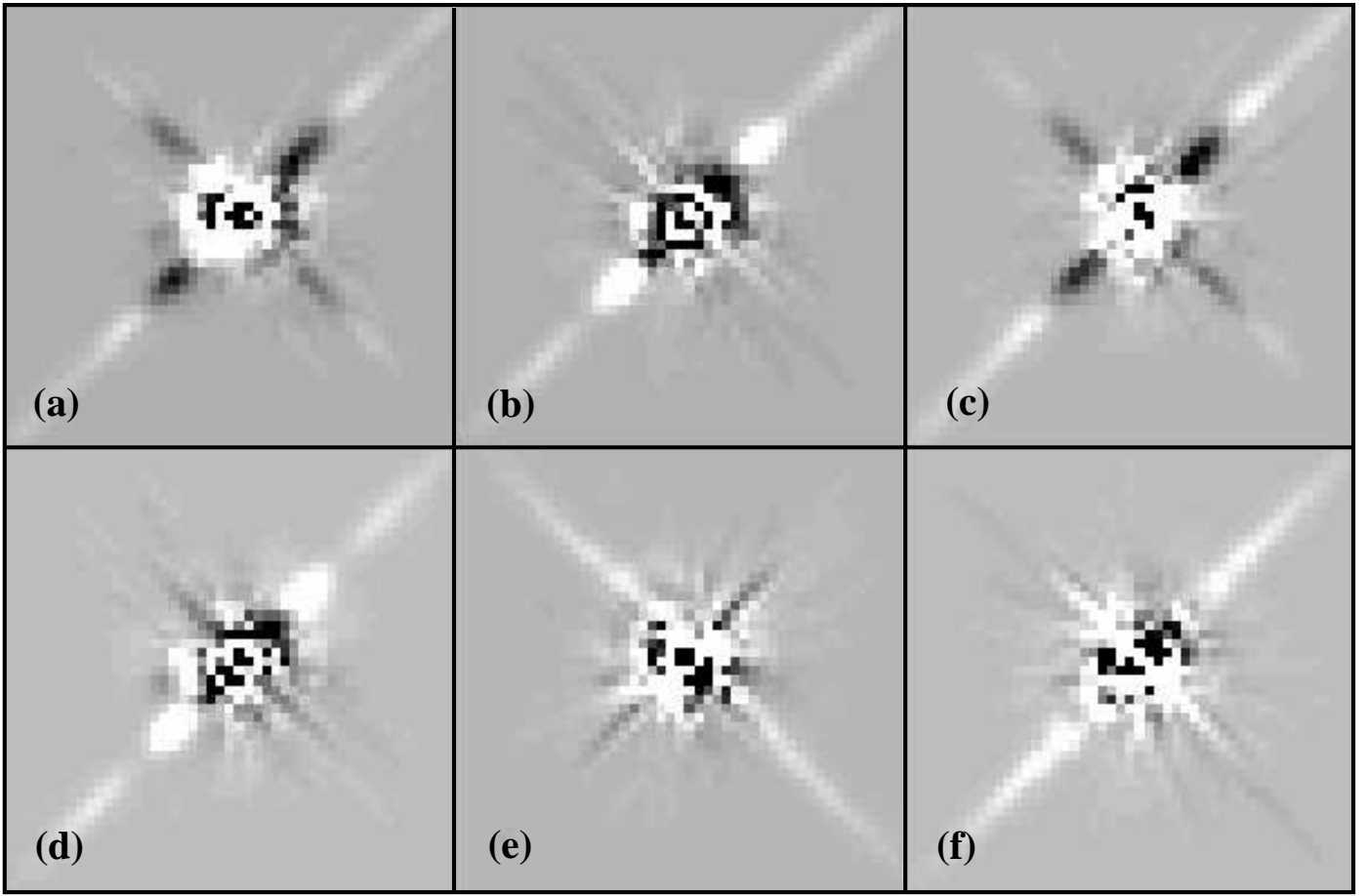,width=18.5cm,angle=0}
\figcaption[fig3.ps]
{
Comparison of the reference PSF with the other PSFs located at different
positions described in Fig. 2. We show the 2-D residual images
after fitting each PSF to the reference PSF.  The pixel offsets with respect
to the reference PSF are ({\it a}) (0,100), ({\it b}) ($-$100,0), ({\it c})
($-$100,$-$100), ({\it d}) (100,100), ({\it e}) ($-$300,300), and ({\it f})
($-$300,$-$300). The size of each box is 50$\times$50 pixels
($\sim 5$\asec$\times$5\asec).  All images are on a linear gray scale.
The PSF variation due to position changes is
substantial.
\label{fig3}}
\end{figure*}

\begin{figure*}
\psfig{file=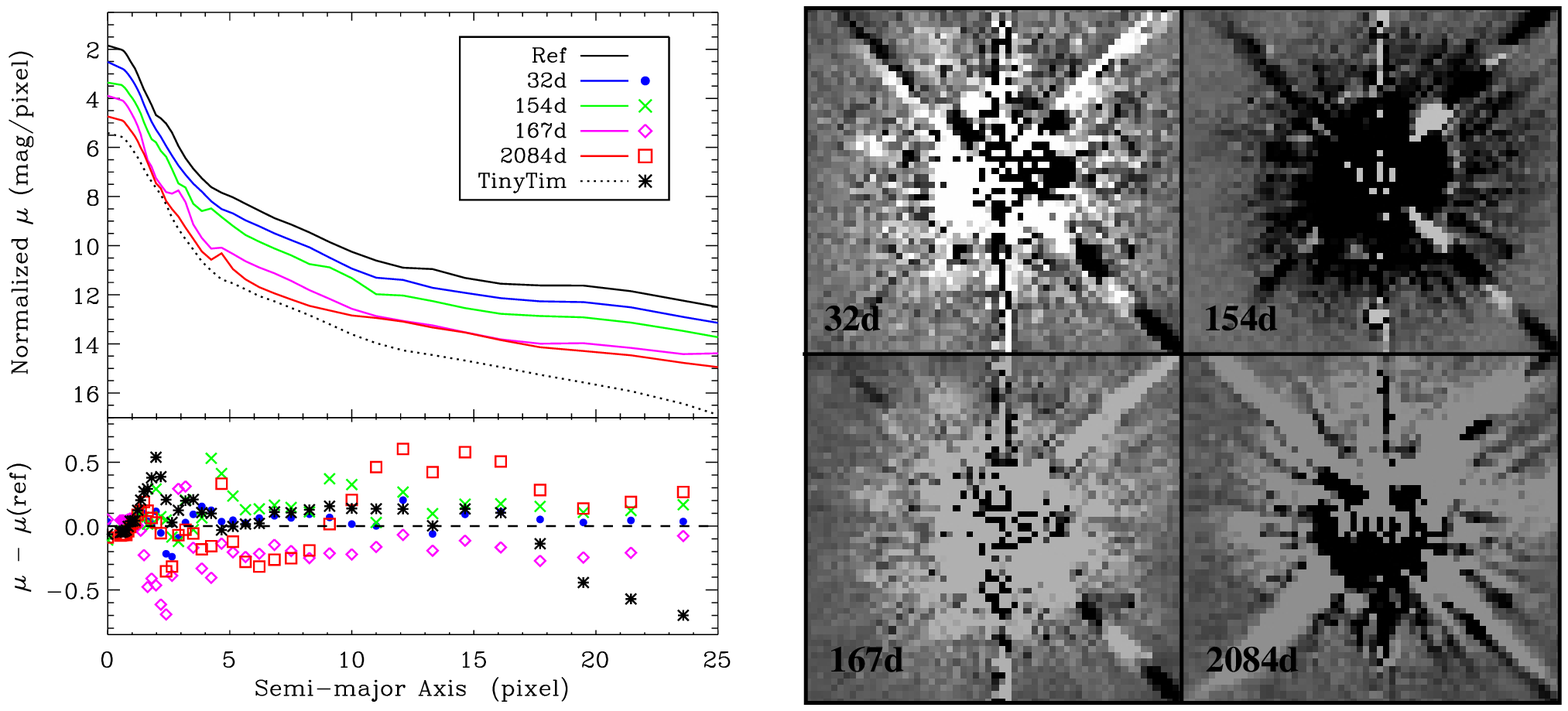,width=18.5cm,angle=0}
\figcaption[fig4.ps]
{
({\it Left}) Similar to Fig. 1, except that here we show PSF variation as a
function of time.  The reference PSF ({\it black line}) was
observed on 1995 October 1 with the WF3 detector and the F606W filter.  The
other PSFs were observed with the same filter and detector 32 days
({\it blue line and circles}), 154 days ({\it green line and
crosses}), 167 days ({\it magenta line and diamonds}), and 2084 days
({\it red line and squares}) later. We also compare another PSF made with
TinyTim ({\it dotted line and asterisks}).
({\it Right}) Residual images resulting from fitting each PSF with the
reference PSF. The size of each box is 60$\times$60 pixels.  These results show
that temporal variations of the PSF are larger than variations due to spectral
changes. All images are on a linear gray scale.
\label{fig4}}
\end{figure*}

Figure 2 shows that the degree of variation grows with
increasing distance from the center of the chip.  Spatial variations, at
distances separated by less than 100 pixels, affect the wings, but not much
the core.  In fact, the core appears stable to within 10\%, whereas the wings
can vary by up to 30\%.  On the other hand, for PSFs separated by more than 100
pixels, the differences are large both in the wings and in the core, by as
much by 30\%--50\%.  Figure 3 visually shows the residuals between each of the
PSFs relative to the reference.  In light of the systematics observed in
Figure 2, it is clear that even if no host galaxy light is detected beneath an
AGN, sometimes it is possible to misconstrue its presence, especially in 1-D
surface brightness profiles.  In 2-D it can be seen that much of the excess
flux in the wings at larger radii is due to an imperfect match in the
diffraction spikes, which make up a high fraction of the flux locally.  From
this experiment we conclude that to minimize PSF systematics, it is crucial to
find PSF stars that are observed to well within 100 pixels from the reference
position of the science target.

{\it Temporal variability} \ \ \ \ \ To quantify temporal variability, the
PSFs need to be observed in different orbits, so TinyTim is not suitable for
this experiment.  To obtain PSFs separated in time, we searched in the \hst\ 
Archive
for stars observed with the same detector (WF3) and same filter (F606W), but
in different orbits.  Our requirement is that they have sufficiently high 
signal-to-noise ratio ($S/N$) in the wings while being not so bright that they 
are saturated in the core, and that they are located close to the same chip 
locations.  There are only five stellar images that meet these criteria in the 
\hst\ Archive.  And, while four of the PSF stars were observed in the same 
\hst\ 
program and located at the same position, the fifth was placed 90 pixels away.
Furthermore, while the observations were done within 5 months for four images, 
the last one was obtained 6 years later. Therefore, we can test both the short- 
and the long-timescale variations with these five PSF stars.

We fit all the stars to the reference star, defined here to be that observed 
at the earliest epoch. We subtracted the sky value for all images before the 
fit was done. The fit, performed in a region of size $60 \times 60$ pixels, 
contains two 
free parameters, namely the position and the magnitude of the star.  To 
achieve a PSF image with high dynamic range, we combined a short exposure of 
the unsaturated core with a long exposure that achieves good $S/N$ in the 
wings.  This procedure still leaves some bleeding columns from the saturated 
core.  The bleeding regions, however, cover only $\sim 1\%$ of the fitting 
region, and the amplitude of the corrupted pixels is less than $10^{-5}$ (12.5 
mag) of the maximum value of the stellar core. Thus, these regions have no 
quantitative effect on the fit.  

Figure 4 shows the 1-D surface brightness profiles and the residual images.  
In comparison to spatial variations and chromatic issues, the temporal 
variation is significantly larger and affects the entire PSF, in both the core 
and the wings.  The variation on short timescales ($\sim$ 1 month, {\it
circles} in Fig.~4) is less than $\pm$0.2 mag, which is slightly less than 
that in 5 months ({\it crosses} and {\it diamonds}). However, the star 
observed 6 years apart ({\it squares}) deviates from the reference star by 
$\pm 0.6$ mag in the full range.  
Comparing also the TinyTim PSF ({\it asterisks}), it is 
obvious that TinyTim does not do well at modeling either the core or the 
wings; in particular, it underestimates the flux in the wings.  The residual 
images from the 2-D fit also show that the profile shape systematically 
changes with time. We note that the PSF star observed $\sim$6 years apart is 
affected not only by temporal variations but also by spatial variation, being 
separated by $\sim$90 pixels from the reference star.

\subsection{Undersampling of the PSF}

In addition to the considerations above, an important issue to bear in mind is
that most of images obtained using \hst\ are undersampled from optical
wavelengths up to $\sim 1.6\, \mu$m (when observed with the NIC2 camera on 
NICMOS).  The effect of undersampling is most severe for the WF CCDs on 
WFPC2.  Undersampling of the PSF is a problem in high-contrast imaging because 
fitting the AGN requires matching its centroid to that of PSF image, which 
involves subpixel interpolation to reassign flux across pixel boundaries.  
When a PSF is not Nyquist-sampled, the interpolation fundamentally does not 
have a unique solution, so that subtracting one unresolved source from another 
introduces a large amount of numerical noise,
which severely hampers high-contrast imaging.  Lastly, image
convolution cannot be performed correctly because of frequency-aliasing in
taking the Fourier transforms of the PSF and the host galaxy model.

In the WF camera of WFPC2, the full width at half maximum (FWHM) of the PSF is 
undersampled at 1.5 pixels in $V$.  Therefore, it is not possible to preserve 
the original shape of the PSF when shifting by a fraction of a pixel.
The program GALFIT uses a tapered sinc + bicubic kernel
in order to maintain the intrinsic width of the PSF under subpixel
interpolation.  This method is used because it preserves flux and is
significantly superior to linear and spline interpolation in regimes
especially near Nyquist-sampling frequencies.  However, when an image is 
undersampled, complications can arise because subpixel interpolation can
significantly change both the amplitude {\it and}\ the width of the unresolved
flux, while the wings of the PSF, being much better sampled, do not change
much.  This problem exists independently of, and affects all, the
interpolation techniques.  The changing of the core-to-wing ratio of the PSF
then becomes numerically degenerate with the profile of the host galaxy
component beneath.

To illustrate the problem of subpixel shifting qualitatively, we shifted two
empirical PSF images by a subpixel unit, here arbitrarily chosen to be 0.3 
pixels.  As 
\begin{figure*}
\psfig{file=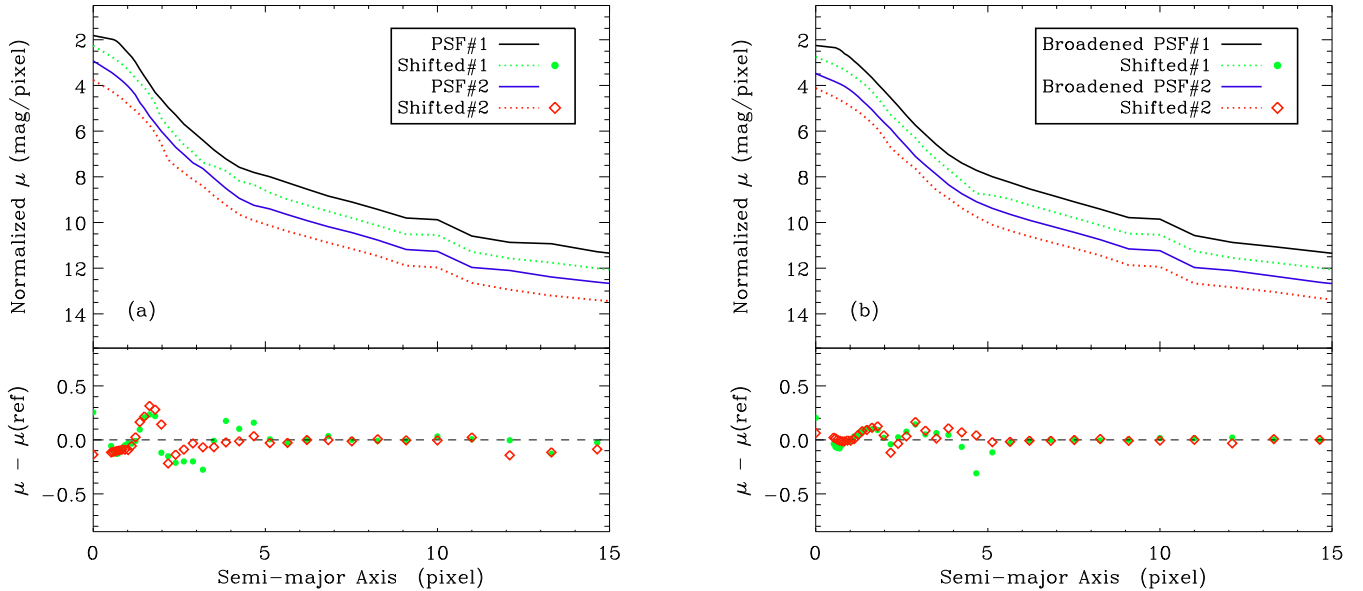,width=18.5cm,angle=0}
\figcaption[fig5.ps]
{
Variation of the PSF from subpixel shifts of the undersampled PSF.  Two
PSFs are shown, each arbitrarily shifted by 0.3 pixels. ({\it a}) Difference
between the original PSF and the shifted one for two observed PSFs.  Large
variations are seen in the central region.  The top panel shows the normalized
surface brightness profiles (each offset by 0.7 units in the ordinate for
clarity) as a function of semi-major axis.  ({\it b}) We broaden the PSFs
using a Gaussian kernel with a FWHM of 2 pixels.  The effect of the subpixel
shift is now significantly reduced.
\label{fig5}}
\end{figure*}
shown in Figure 5{\it a}, while both the width and the flux in the wings do not 
change much, the central value changes by 20\%--30\%.  In the outskirts 
($>$10 pixels), small differences remain due to the fact that the diffraction 
spikes are also not Nyquist-sampled.  Figure 5{\it b}\ shows the same PSFs 
when they are broadened to Nyquist sampling.  Small differences necessarily 
remain because the PSF images are shifted first before being broadened, but 
the agreement is much better.  While superficially there appears to be a 
resolution loss by broadening the PSF in such a way, fundamentally, according 
to sampling theory, there is no more information content in Figure 5{\it a}\ 
than there is in Figure 5{\it b}\ if they are used or analyzed as {\it single}\
images.  Information content is there to achieve ``super''-resolution only 
when subpixel dithering is used to construct the final images.  Figure 
5{\it a}\ indicates that even if the PSF image is taken during the same orbit 
as the host galaxy image, additional errors could be introduced by subpixel 
interpolation no matter how sophisticated is the method of interpolation.  
These figures suggest a way to alleviate the problem, which will be discussed 
in Section 3.4.

\bigskip

In summary, the shape of the PSF is sensitive to changes that occur 
in time and dependent on the position on the detector, but to a much lesser
degree on the SED of the source.  More importantly, in terms of temporal
variability, these results, with caveats about small number statistics,
suggest that even though the telescope cameras undergo refocusing from time to
time, cumulative effects build up through time that may not be fully removed
by refocusing of the camera alone.

For the remainder of the discussion on the simulations, we will pick two 
observed PSF stars: the reference PSF star and another that is taken 1 month 
later and located within 100 pixels of the reference.  For these two PSF stars,
the difference near the peak is 15\% of the maximum, and while systematic 
differences in the residuals are small, the variance in the central 50 pixels 
is 4\% of the peak value, as seen in Figure~4.  Later on, we will also fit the 
simulations with a TinyTim PSF, to contrast with the results obtained through 
using observed PSF stars.

\section{SIMULATIONS}

In the previous section, we identified the factors that most influence the
shape of the PSF.  We find that it is possible to reduce the systematic
differences of the PSFs if certain precautions are taken during the
observational stage.  Nevertheless, even under the most extreme care, 
systematic differences will still propagate into the analysis due to 
the aforementioned effects.  For realistic \hst\ observations, it is often
impractical to observe PSF stars ideal for all the science data in hand.  
Therefore, in this section, we seek to quantify the extent that observations 
are typically affected by performing image-fitting simulations under 
{\it realistic}\ observational conditions, to the extent that it is possible 
to mitigate the PSF mismatches a priori.  We use GALFIT (Peng et al. 2002) 
to conduct the simulations.  GALFIT is designed to perform 2-D profile fitting 
and allows us to fit a galaxy image with multiple components convolved with a 
PSF.  Although other 2-D fitting programs exist (e.g., McLure et al. 1999; 
Schade et al. 2000), the results of this study are sufficiently general that 
the main conclusions should be broadly applicable to AGN imaging studies 
in general.

\hst\ observations of AGNs have been obtained using various exposure times, 
sensitivities, instruments, and filters.  The simulations below are intended 
to be broadly useful for a wide range of past studies using \hst\ as well as 
future studies, where the image sampling is close to Nyquist sampling.  For 
this reason, we adopt the following strategy in producing the simulations:  
the simulations are first and foremost referenced to the $S/N$ of the AGN 
nucleus (defined below). The $S/N$ of the unresolved nucleus is often the 
easiest parameter to measure in luminous AGNs, and is more general than the 
luminosity 
parameter.  Secondly, at a given \nsn\ in the simulation, the host galaxy flux 
normalization is referenced to that of the AGN---a quantity that we call the 
``luminosity contrast'' (\hnr)---instead of using the absolute luminosity of 
the galaxy.  Luminosity contrast is often the factor that determines the 
reliability of host galaxy detection:  when \hnr\ is low, it is hard to detect 
the host galaxy, whereas when \hnr\ is high, the AGN is harder to measure.  
Both low and high \hnr\ are considered to be high contrast.  The strategy of 
using \nsn\ and \hnr\ makes irrelevant information about the exposure time and 
intrinsic luminosity when quantifying parameter uncertainties of the AGN and 
the host.  Therefore, once the \nsn\ and \hnr\ parameters are 
measured, uncertainties in the measurement parameters can be estimated by 
referring to the appropriate figure for a given \nsn.

The $S/N$ of the unresolved nucleus is calculated, without loss of generality, 
by defining that all the flux is concentrated within a single pixel.  Thus,

\begin{equation}
(S/N)_{\rm nuc}= \frac{F_{\rm nuc}}{\sqrt{F_{\rm nuc}+Rd^2+Sky+Dark}},
\end{equation}

\noindent where $F_{\rm nuc}$ is the total flux of the unresolved AGN nucleus, 
$Rd$ is the readout noise, $Sky$ is the sky value, $Dark$ is the dark current, 
and all quantities are in units of electrons.  The $S/N$ for the nucleus 
calculated in this manner is slightly larger than the value obtained from that 
accounting for the fact that the point source has a finite width.  However, 
since the unresolved core is not in the read-noise or sky-dominated regime, 
and most of the
flux is contained within 1 to 2 pixels, the noise of the core hardly depends on 
the exact area. The advantage of our definition is that it can be easily 
renormalized to other areas in lower $S/N$ regimes.  We note that the main 
results of this paper are actually not very sensitive to the exact value of 
\nsn.

\subsection{Creation of the Simulation}

The reference simulation models were produced using a single reference PSF to 
represent the AGN nucleus and for image convolution.  The light profile of the 
host galaxy was parameterized as a single-component \ser\ (1968) profile,
convolved with the reference PSF.  We made 10,000 reference images, by varying
the following parameters:  (1) $S/N$ of the nucleus [$50 \leq (S/N)_{\rm nuc} 
\leq 5000$] on a logarithmic scale, (2) luminosity ratio between galaxy and 
nucleus ($0.01 \leq L_{\rm host}/L_{\rm nuc} \leq 100$) on a logarithmic 
scale, (3) effective radius ($ 1 \ {\rm pixel} \leq R_{e} \leq 50 \ {\rm 
pixel}$) on a logarithmic scale, (4) axis ratio ($0.7 \leq b/a\leq 1$) on a 
linear scale, and (5) \ser\ index ($n=1$, equivalent to an exponential 
profile, {\rm or} $n=4$, equivalent to a de~Vaucouleurs profile).  The values 
were randomly selected in these ranges.  The range of values for the host 
galaxy parameters ($R_{e}$, $b/a$, and $n$) was determined by the typical 
range seen in actual observations. The size of each simulated image is 
$\sim$10 times larger than $R_e$.  To maximize efficiency for the simulations, 
we use a convolution size of $50 \times 50$ pixels for the PSF image; we 
verified that enlarging the convolution size to $100 \times 100$ pixels has no 
noticeable impact on the results.

To cast the simulations into more concrete terms, the foregoing parameters,
translated to single-orbit, 3000-second \hst\ images, mean that the range of 
\nsn\ corresponds to $15 \leq m_{\rm F606W} \leq 25$, with the host 
galaxies spanning $\pm 5$ magnitudes around that range.  This sufficiently 
covers most of the data contained in the \hst\ archive, and the dynamic range 
is large enough for most practical observations. 

The simulations below examine the most typical circumstances encountered in
high-contrast images.  During the fit, we allow the following sets of 
parameters to be 
free: (1) position, luminosity, effective radius, axis ratio, position angle, 
and S\'{e}rsic index for the galaxy; (2) position and luminosity for the 
nucleus; and (3) sky value.

\begin{figure*}
\psfig{file=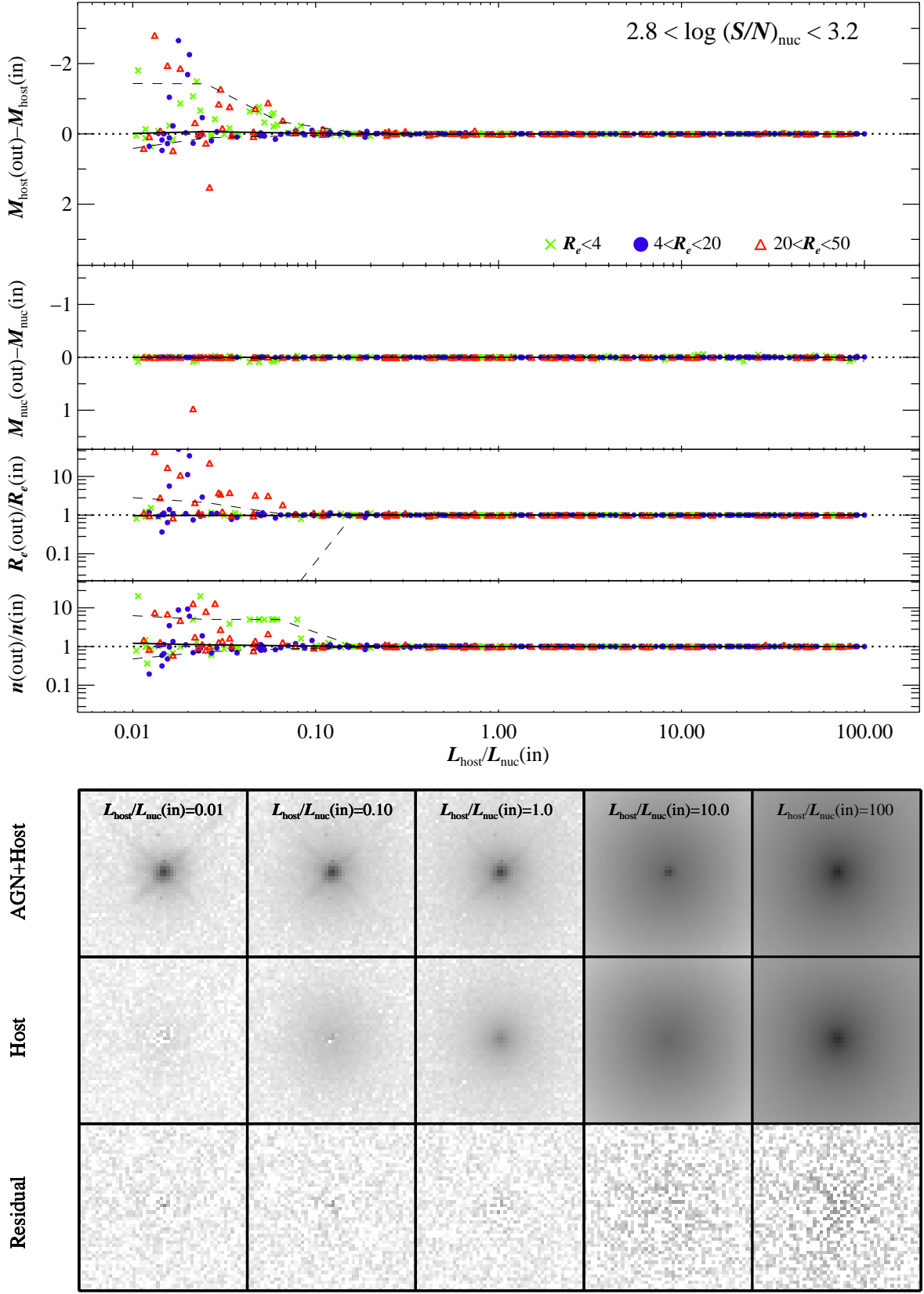,width=17cm,angle=0}
\figcaption[fig6.ps]
{
Simulation results under idealized conditions. We fit artificial images with
the same PSF as that used for generating the input images. The upper part of
the figure shows residuals in the magnitude of the host ($M_{\rm host}$),
magnitude of the nucleus ($M_{\rm nuc}$), effective radius ($R_e$), and \ser\
index ($n$) as a function of input luminosity ratio between the host and the
nucleus (\hnr) and input signal-to-noise ratio of the nucleus [\nsn].  Host
galaxies of different effective radii are marked in ({\it green}, $R_e < 4$
pixel), ({\it blue}, $4 < R_e < 20$ pixel), and ({\it red}, $20 < R_e < 50$
pixel).  Here, we only
show the results for $10^{2.8} \leq$ \nsn\ $\leq 10^{3.2}$, which is typical
of most actual objects observed with {\it HST}.  The plots for other values
of \nsn\ are presented in the Appendix.
We plot only 25\% of the data points to avoid crowding; this subset of
points adequately represents the overall trends.
The solid lines represent the median
value for the simulated sample, while the dashed lines demarcate the region
that encloses the central 70\% of the sample.  The lower portion of the figure
shows examples of the artificial and residual images as a function of
increasing \hnr\ (from left to right). All images are on an asinh stretch.
\label{fig6}}
\end{figure*}
\begin{figure*}
\psfig{file=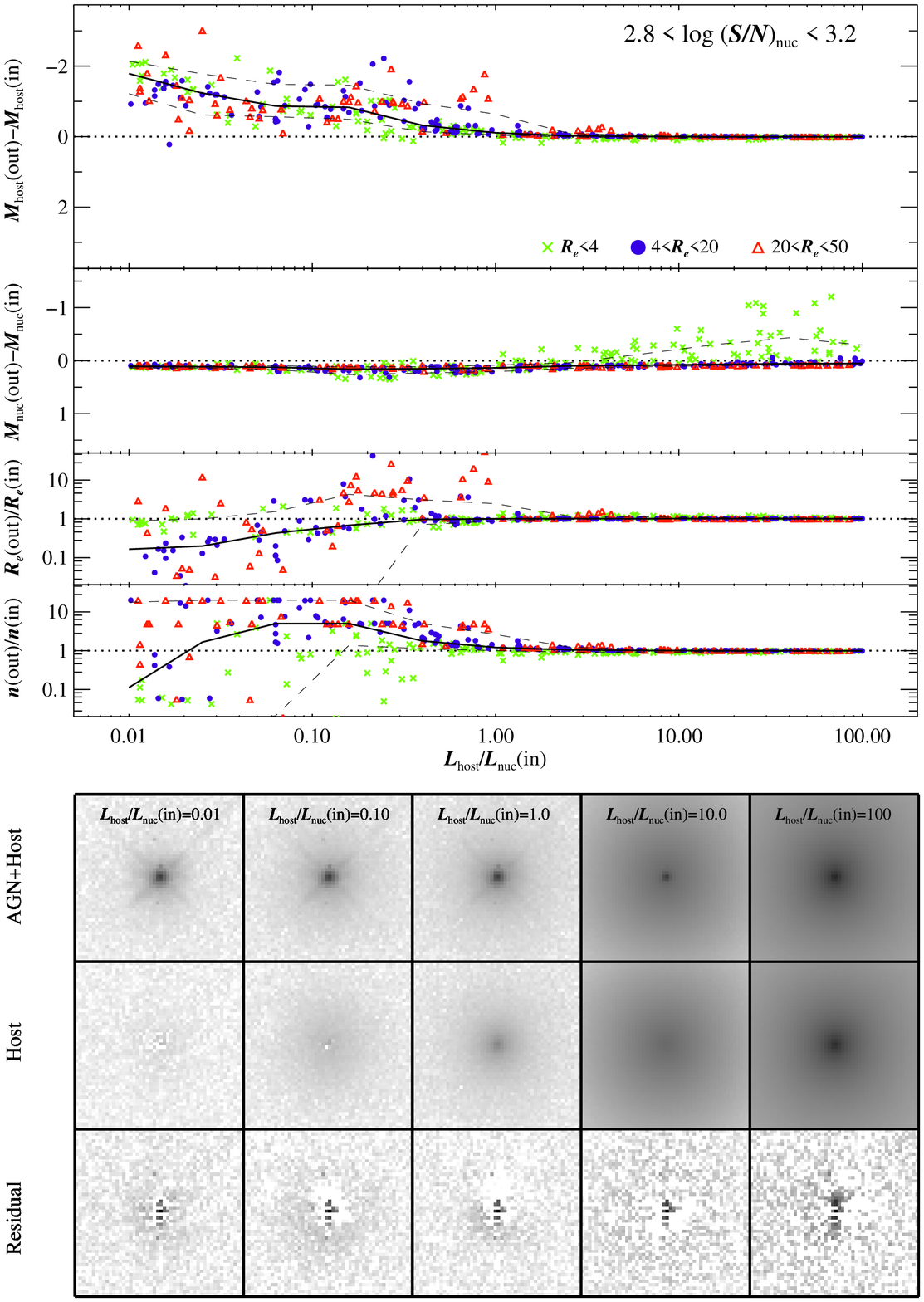,width=18cm,angle=0}
\figcaption[fig7.ps]
{Similar to Fig. 6, except that here we fit the artificial image with a PSF 
different from the one used for generating the input image.  The conditions of 
the simulation are more realistic than those in Fig. 6. This test 
shows that PSF mismatch is the main culprit for the systematic uncertainty. 
\label{fig7}}
\end{figure*}
\begin{figure*}
\psfig{file=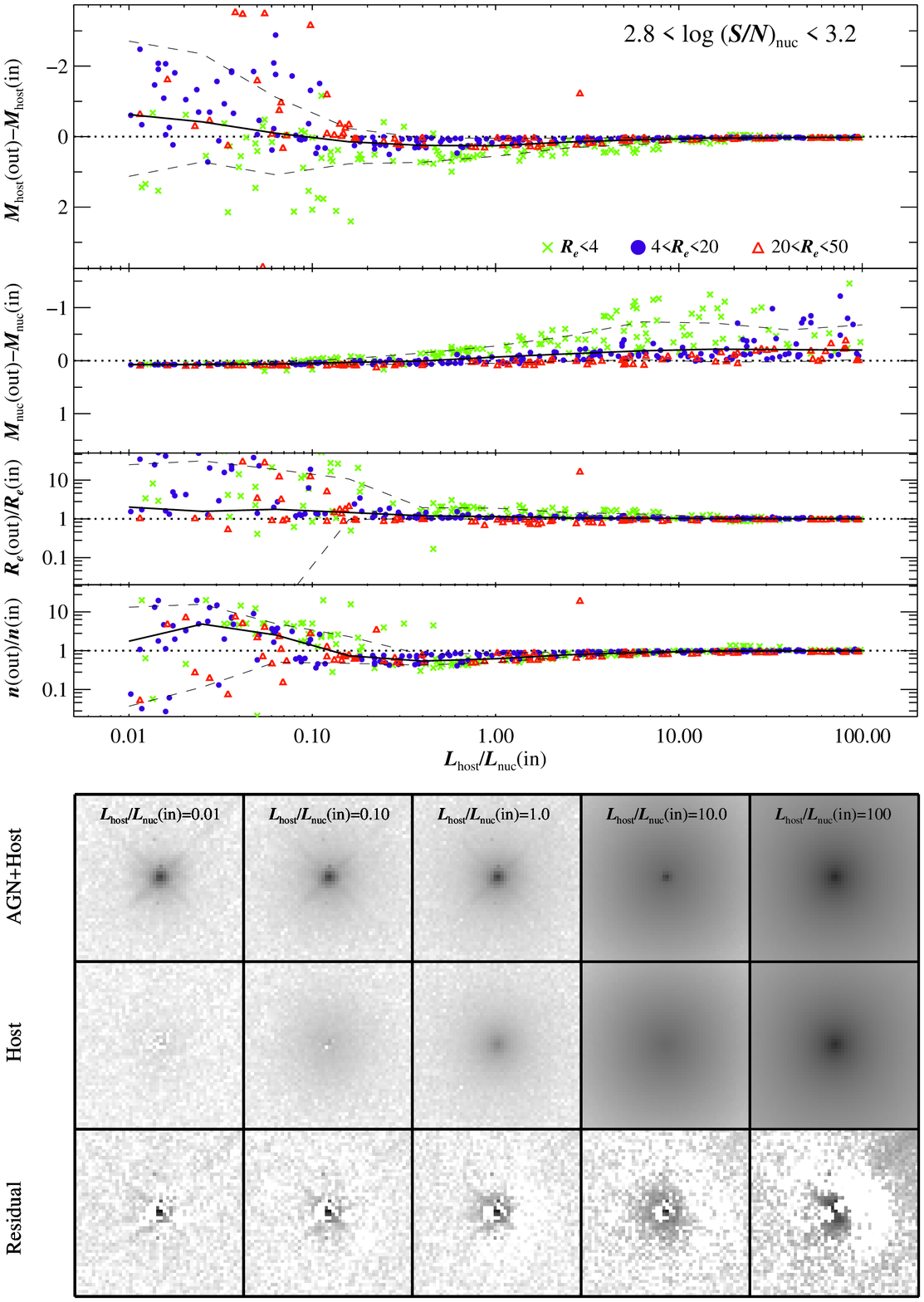,width=18cm,angle=0}
\figcaption[fig8.ps]
{Similar to Fig. 6, except that here the fits were done with the TinyTim PSF, 
which was not oversampled. The conditions of the simulations are more 
realistic than those in Fig. 6.  This test suggests that the scatter for the 
host parameters is due to PSF mismatch, but the systematics, except for the 
nuclear magnitude, seem to be better controlled than in Fig. 7.
\label{fig8}}
\end{figure*}
\subsection{Idealized Simulations}

As a point of reference, our first set of simulations is designed to be highly
idealized: we fit the artificial images with the same PSF used to create them.
Although this simulation does not account for PSF mismatch, it gives us a
``zero point'' expectation for how well we can extract the host galaxy
parameters in the photon limit. The simulation results are summarized in
Figure 6, where we show the residual of galaxy luminosity, nucleus luminosity,
effective radius, and \ser\ index as a function of the luminosity contrast.
The \nsn\ of $\sim 1000$ in Figure 6 corresponds to a point source magnitude
of 18.5 mag in the F606W filter, for a single \hst\ orbit.  In the 
Appendix, we show plots of other \nsn\ regimes. 
In Figure 6, when the luminosity of the AGN is more than 10 times the host
luminosity, the underlying galaxy tends to be dominated by the noise of the
nucleus.  In this case, GALFIT tries to extract the host galaxy component from
the nucleus itself, which leads to the \ser\ component having a smaller size
and a higher \ser\ index.  The size of the galaxy is also important. When $R_e$ 
is small ({\it green dots}) and difficult to distinguish from the
nucleus, or large ({\it red dots}) and has low surface brightness,
the scatter in all morphology parameters is large.  However, we find that the
scatter is not very dependent upon the \ser\ index of the host galaxy.
Most importantly, when the AGN is brighter than 20:1, the host galaxy
detection becomes quite difficult even under ideal conditions.  This is
similar to findings in nearly all quasar host galaxy studies where the hosts
are often not much fainter than 3 magnitudes compared to the AGN.  Under
certain situations it is possible to detect quasar hosts that have a higher
contrast with the AGN, such as when the host has a high ellipticity (and thus 
a high surface brightness).  In these limits, the fluxes of even real host
detections are likely to be biased systematically high.

\subsection{Testing PSF Mismatches}

To test PSF mismatches we fit the simulated data with a different PSF from
the one that was used to
create them.  We first test the results by fitting with a stellar PSF, and
next with a model PSF generated by TinyTim.

{\it Stellar PSF.}\ \ \ \ \ In this simulation the analysis PSF (FWHM $\approx
1.6$ pixels) was observed one month apart from that (FWHM $\approx1.5$ pixels) 
used to generate the 10,000 reference data models.  This experiment is more 
realistic because most AGN observations do not have concurrently observed 
PSFs; they rely instead on PSFs in the archive or TinyTim models.  

The simulations (Fig. 7) show that even when the PSF matches are fairly
good, as seen in the residual images, the systematic errors can be quite large.
When the contrast is high, the effective radius and \ser\ index plots show
that GALFIT may try to reduce the PSF mismatch by using the host galaxy \ser\
component, the result of which is to push the concentration index to either
extreme and to make the galaxy size small.  The host galaxy luminosity robs
light from the quasar itself in the process.  For compact galaxies, the
systematic errors start to increase around \hnr $\approx 1$.  For low-surface 
brightness galaxies ($R_{e} > 20$ pixels), the errors occur at higher \hnr.
We note that GALFIT does not permit \ser\ indices larger than $20$, 
which is why the ratio of the \ser\ indices saturates at a value of 5 [when 
$n$(in) = 4] or 20 [when $n$(in) = 1].

We further find that the systematic trend is slightly dependent on the \ser\ 
index, in the sense that the hosts with $n$(in) = 1 is better recovered than 
those with $n$(in) = 4, but less so when \nsn\ \gax\ 250. When the host has 
an $n=4$ profile, the recovered flux from the host can be easily buried under 
the structure of the PSF mismatch.  Thus, it is hard to accurately detect an 
AGN host with a high \ser\ index (central concentration) and small size. At 
moderate to high \nsn, there is only a small dependence on \nsn\ because the 
systematics are dominated by the PSF mismatch. 

As seen in Figure 7, there is a clear offset ($\sim 0.1 $ mag) in the
luminosity of the nucleus. As explained above (\S{2.2}), and as shown later in
\S{3.4}, this effect is caused by the undersampling of the PSF, so that the
cores cannot be adequately sampled and modeled.  This then leads to an
overestimate or underestimate of the halo of the PSF, which results in 
erroneous inferences on the host galaxy properties when the galaxy is faint. 
In our case, the underestimate of the nuclear flux boosts the flux of either 
the sky or the host luminosity.  We discuss later in \S{3.4} how this 
systematic error can be reduced.

\bigskip
{\it TinyTim PSF.}\ \ \ \ \ In addition to fitting the artificial images with
an actually observed PSF star, we also tried using a PSF generated with 
TinyTim (Fig.  8).  When there is no stellar PSF observation, this is an 
often-used strategy in the literature.  We create a 1$\times$ sampled PSF for 
this analysis.  Figure 8
shows that the fit with this synthetic PSF is less accurate compared to the
previous two simulations using an observed star.  In particular, if the size
of the host galaxy is small, the parameter recovery is quite poor, {\it even if
the residuals may not indicate there to be an obvious sign of a PSF mismatch},
especially at low \hnr.  This point cannot be over-emphasized.

Since the TinyTim PSF is sharper than the stellar PSF, the mismatch in the 
central few pixels is more problematic due to undersampling.  As in the previous
simulation, when \hnr\ is large, the AGN flux is measured to be too luminous,
even though the host luminosity is not affected much. Therefore it is hard to
get accurate fits for both the host and the nucleus simultaneously at the
extreme ends of the luminosity contrast if one uses a non-oversampled TinyTim
PSF. 

We also fit the simulated images with a $4\times$ oversampled TinyTim PSF. 
When an oversampled PSF is created by the TinyTim program, it is not
convolved with a charge diffusion kernel. We manually convolved the 
oversampled PSF with the charge diffusion kernel.
We find no significant differences from the fits with a non-oversampled 
TinyTim PSF.

\subsection{Broadening the PSF and Science Images} 

To minimize interpolation errors, we need to model both the gradient and the
curvature of the PSF, which is not possible when the PSF is undersampled.  As 
discussed in \S~2.2, interpolation errors necessarily occur when an 
undersampled PSF is shifted by a fraction of a pixel, regardless of the 
algorithm used in the interpolation.  To see if this crucial problem can be 
reduced, we repeated the same simulations by slightly broadening both the PSF 
star and the simulated images as they are observed.  To 
achieve Nyquist-sampling, we want the final PSF to have a FWHM of 2 pixels.
Since the FWHM of the original PSF is $1.5-1.6$ pixels, we use a 
Gaussian kernel with FWHM = 1.45 pixels for broadening.

As shown in Figures 9 and 10, the results improve considerably.  While the
behavior of the systematic errors on the host galaxy luminosity is consistent
with previous results in Figures 7 and 8, the upturn point (\hnr $\approx 0.5$)
where GALFIT begins to overestimate the host galaxy luminosity is a factor of
2 less than previously (\hnr $\approx 1.0$).  This experiment also recovers 
the luminosity of the nucleus much better than before, even if the host galaxy 
is small.  In addition, the artificial offset in the residual of the AGN
luminosity is virtually eliminated.  Lastly, the scatter in the fitted
parameters has, in all cases, decreased quite substantially.

Interestingly, the improvements for a broadened TinyTim PSF are more dramatic
(compare Fig. 10 with Fig. 8), such that a TinyTim PSF {\it can} sometimes work
better than a real star.  The latter fact is probably just a coincidence and
not likely to be true in general due to the temporal variability of real PSFs.
Nevertheless, the improvement does suggest that if there is no stellar PSF
observed, TinyTim can be an acceptable substitute, if both the PSF and the
data images are oversampled in the same way.  Comparing Figure 10 with 
Figure 9, in which the fit was done with the broadened stellar PSF, we can
conclude that the recovery for host magnitude is reasonable, but the AGN
luminosity may still be somewhat biased when the host galaxy is much more
luminous.

In summary, the results of these tests indicate that image decomposition
should be performed only on images that are Nyquist-sampled.  If both the PSF
and the data are not sampled adequately, one can simply convolve both images
with the same Gaussian kernel.  Alternatively, dithering can be used 
to achieve Nyquist sampling when
observing the data and the PSF images.

\subsection{Holding the S\'ersic Parameter Fixed}

When the host galaxies are faint and hard to deblend from their central AGN,
one common technique used in the literature is to hold the \ser\ index $n$ 
fixed, while allowing other parameters to converge (e.g., Jahnke et al.  2004; 
S\'{a}nchez et al.  2004).  We perform similar experiments to test how reliably 
we can extract the host parameters by using such a prior. We create artificial
images for \nsn\ $= 1000$, varying $n$ from 0.8 to 6.8.  We only
run simulations in the regime where \hnr $\approx 0.3$ and $R_{e}=7$ pixels,
as this is the part of the host galaxy parameter space where the systematic
uncertainties are largest.  Fits are performed in four different ways: fixing 
$n$ = $1$, fixing $n$ = $4$, fixing $n$ to the input value, and allowing $n$ 
to be free.

Figure 11 summarizes the results.  For simulated galaxies where the input
\ser\ indices are $n\ge2$, the luminosity can be recovered to better than 0.3
mag by simply holding the fitted profile to $n=4$ ({\it filled blue circles}).
This is considerably better than the results that allow the \ser\ index to be 
free ({\it stars}).  In addition, if the intrinsic \ser\ index of the host is 
$n < 2$, then holding the fit to $n=1$ also produces a result that agrees 
to better than 0.3 mag.  Therefore, like previous studies, we find that by
using a correct prior, the recovery of other host galaxy parameters can
dramatically improve.  The caveat, however, is that the prior may be hard to
determine when the host is not well-resolved.  Nevertheless, as the bimodality
decision is quite coarse, the decision about which prior to make can in some
cases be simply based on selecting the lower of the two $\chi^2$ values.

\begin{figure*}
\psfig{file=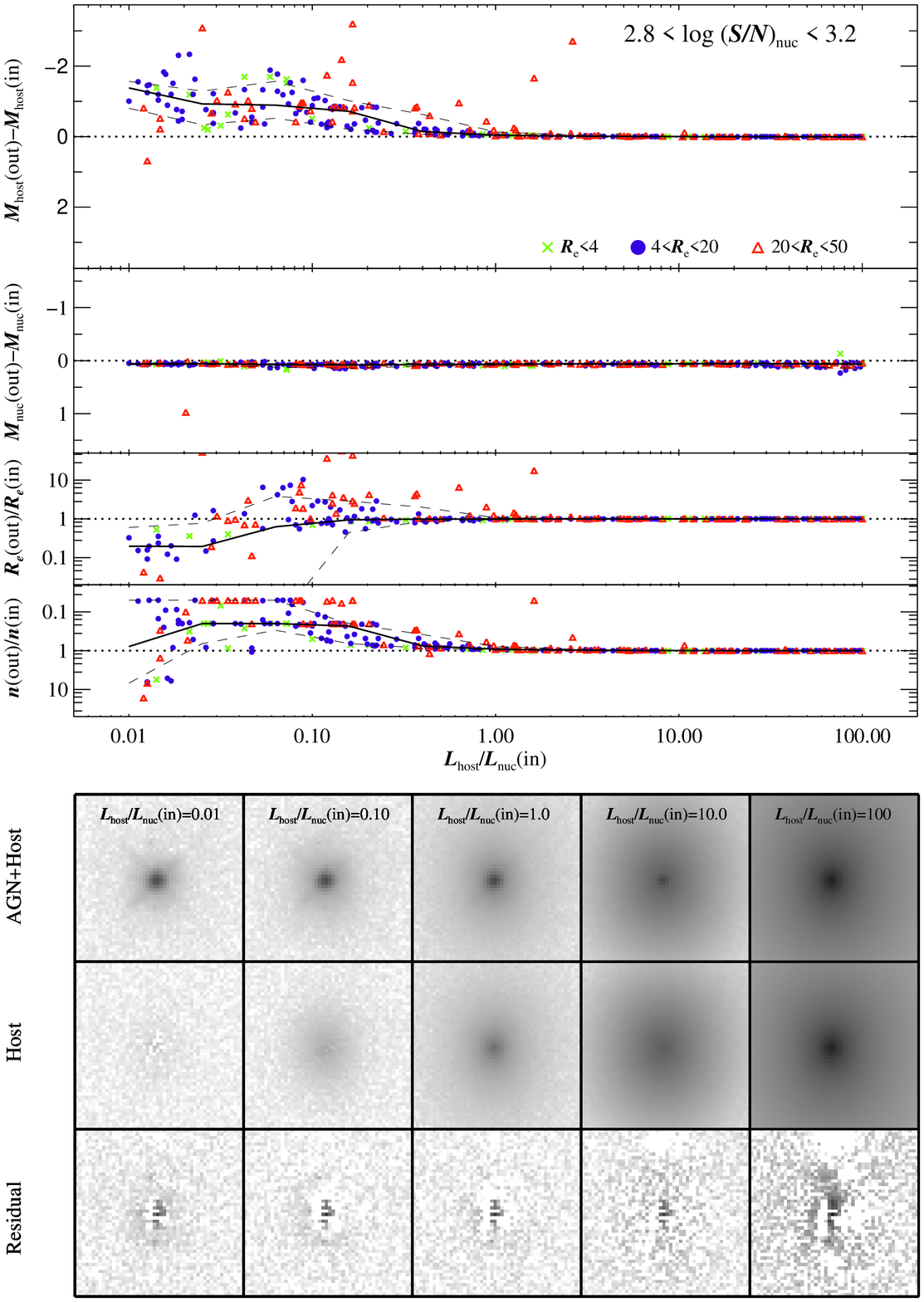,width=18cm,angle=0}
\figcaption[fig9.ps]
{Similar to Fig. 7, except that here we generated the artificial image
with an empirical PSF broadened by a Gaussian with FWHM = 2 pixels, and the 
fits were done with a different, similarly broadened PSF. The input parameters, 
especially the AGN luminosity, are recovered better than in Fig. 7.
\label{fig9}}
\end{figure*}
\begin{figure*}
\psfig{file=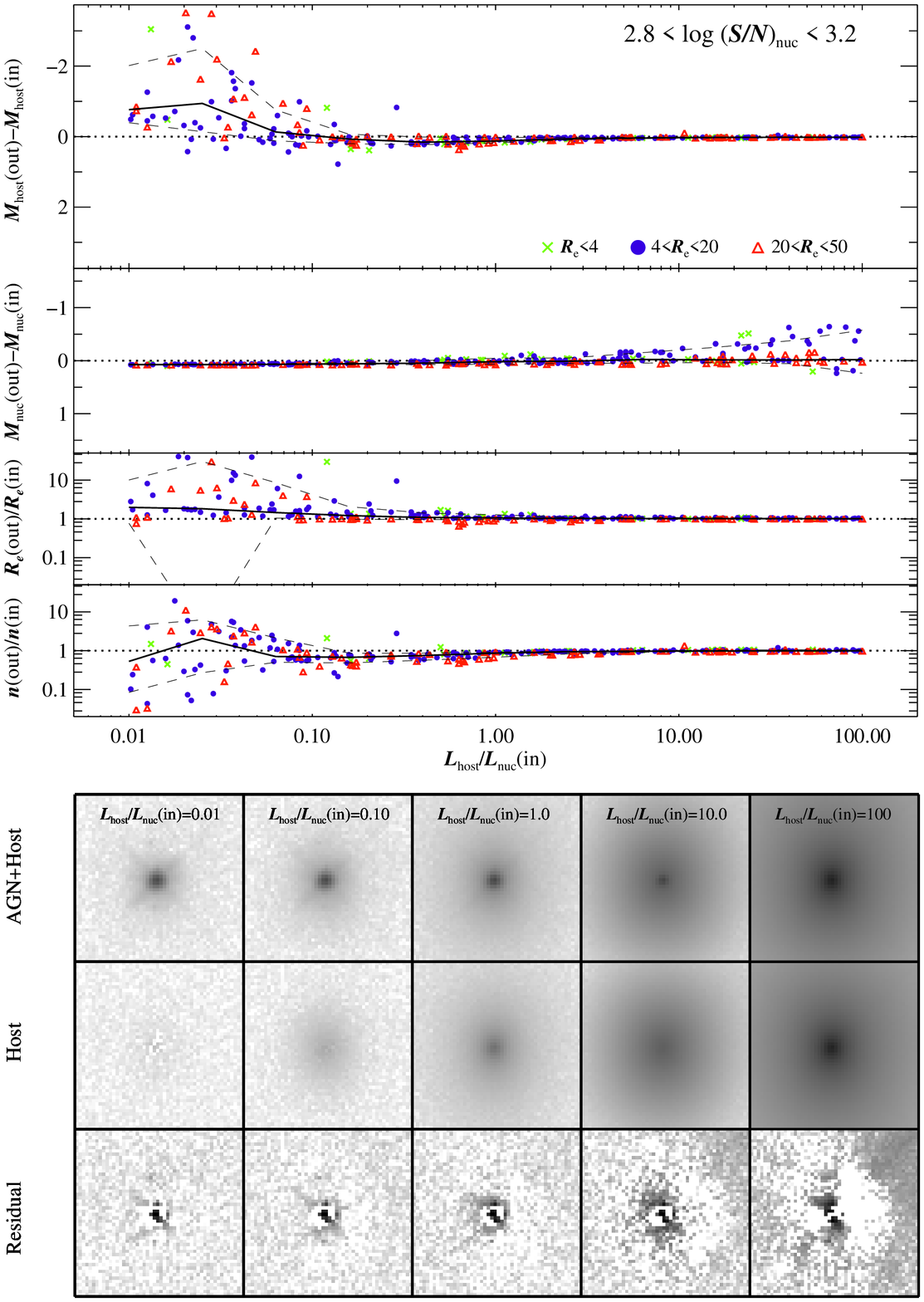,width=18cm,angle=0}
\figcaption[fig10.ps]
{Similar to Fig. 9, except that here we generated the images with a 
broadened empirical PSF and fitted them with a broadened TinyTim PSF.  
The recovery of 
the parameters, especially the AGN luminosity, is substantially improved 
compared to Fig. 8.  
\label{fig10}}
\end{figure*}
\begin{figure*}
\psfig{file=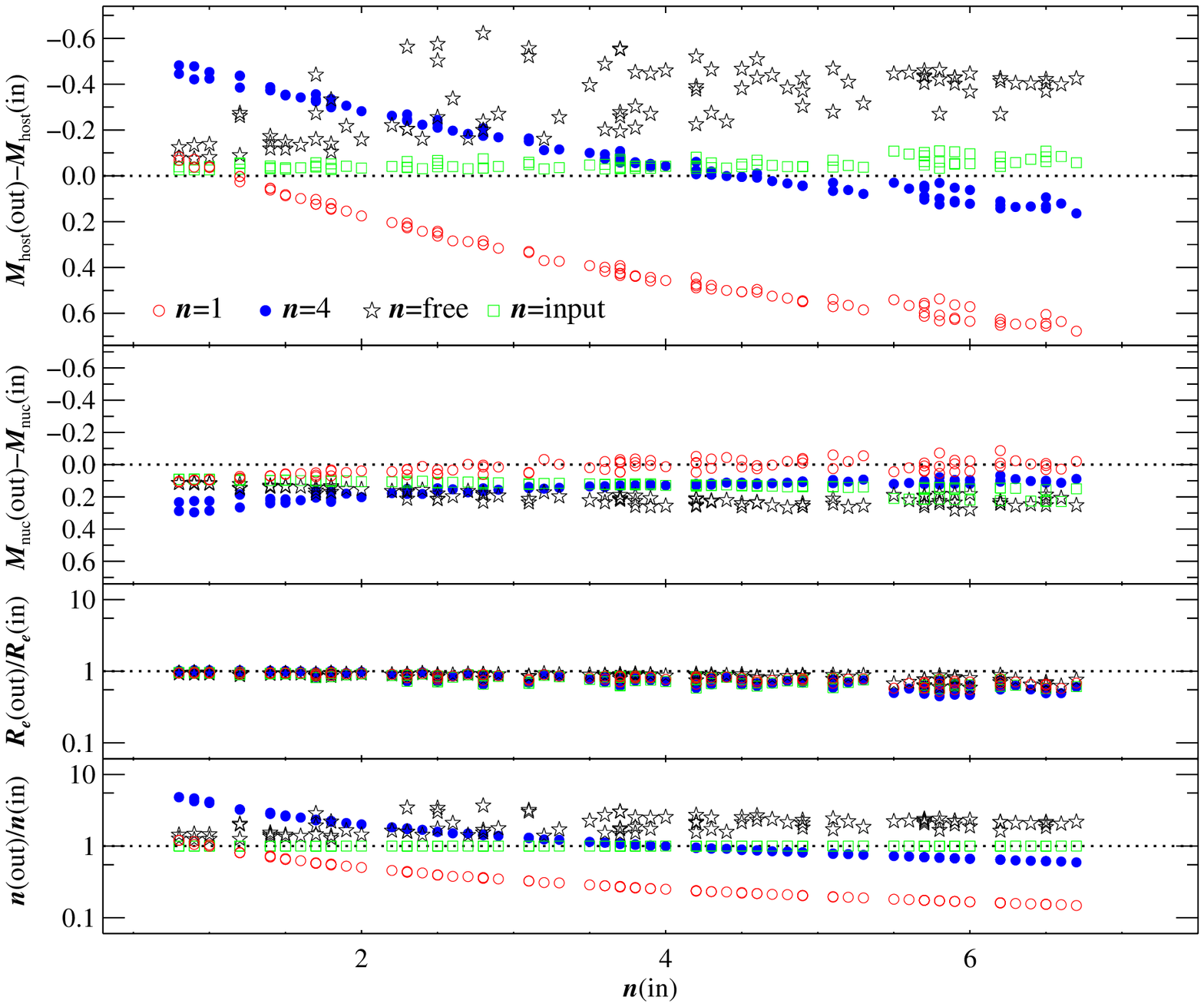,width=18cm,angle=0}
\figcaption[fig11.ps]
{Fitting results with the \ser\ index fixed. After we created artificial 
images with $n$ varying from 0.8 to 6.8, we fit them in four different 
ways: fixing $n$ = 1 ({\it open red circles}), fixing $n$ = 4 ({\it filled blue
circles}), fixing $n$ to the input value ({\it green squares}), and allowing 
$n$ to be free ({\it stars}). For $n>2$, fixing $n$ = 4 better recovers the 
input parameters.
\label{fig11}}
\end{figure*}

\subsection{Reversing the Role of the PSFs}

Previously, our simulation models were created using a single PSF that has 
FWHM $\approx$ 1.5 pixels, but which is then fitted using another with FWHM
$\approx$ 1.6 pixels.  This resulted in systematic biases on the host and AGN
parameters in a certain direction.  A natural follow-up is to reverse the role
of the PSFs to see how the systematic trends may change.  As in Figures
7--10, we only show results for $R_{e} < 20$ pixels and
$10^{2.8}<$\nsn$<10^{3.2}$.  Figure 12 shows the results of this set of
simulations.  We again find that the recovered hosts appear slightly
overluminous, but the systematic errors are smaller than the original
experiment.  Just like before, we find that hosts with $n=1$ and $R_{e} > 4$ 
pixels tend to be recovered well, and the scatter of both the nuclear luminosity
and host luminosity are larger when the hosts are small.  Even though the
degree of systematics is slightly changed, reversing the roles of the PSFs
does not affect the systematic error offsets in exactly the opposite
direction.

\begin{figure*}
\psfig{file=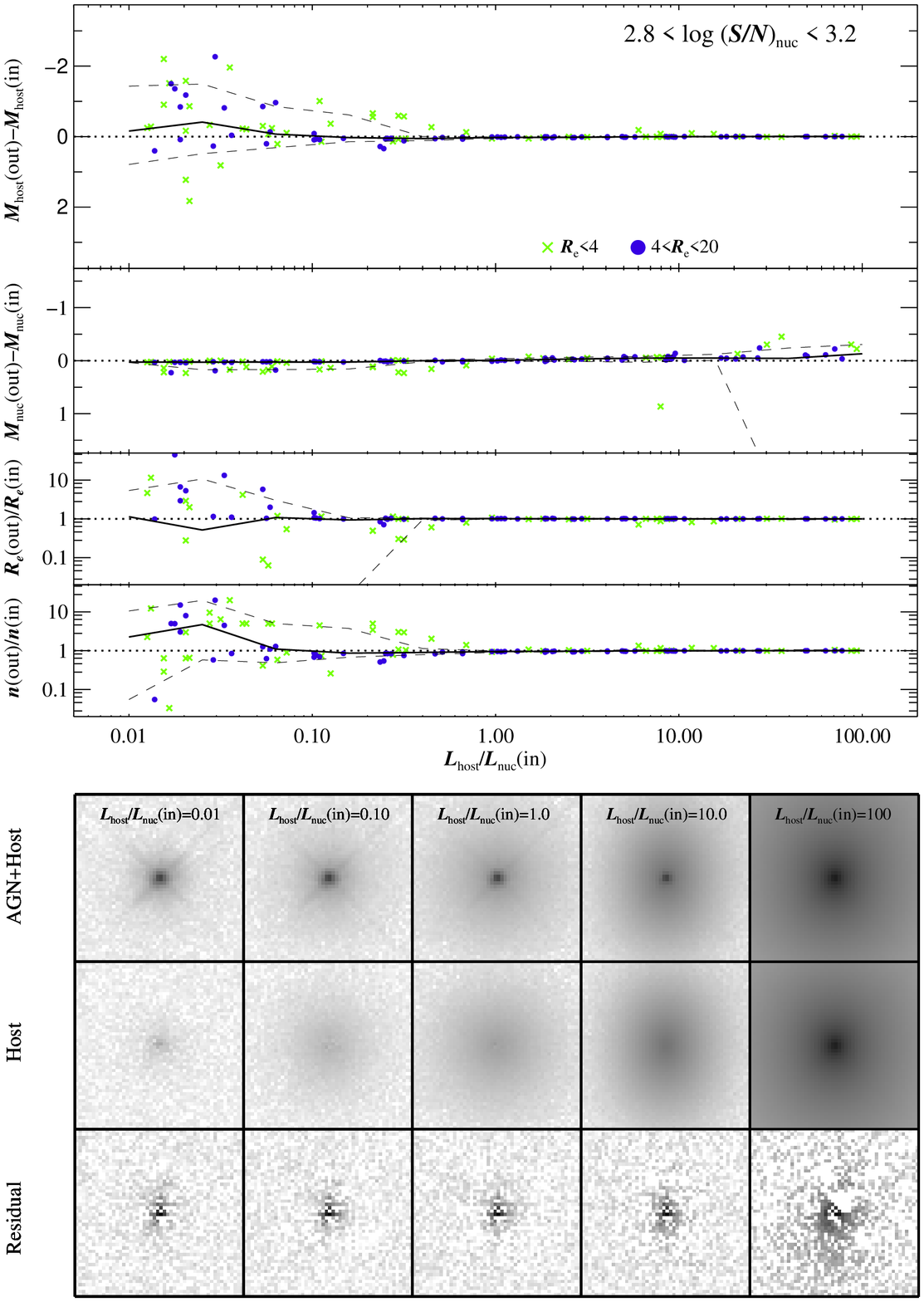,width=18cm,angle=0}
\figcaption[fig12.ps]
{Similar to Fig. 7, except that here we switched the role of the two PSFs used 
for generating the artificial image and for doing the fit. For simplicity, we 
only simulate galaxies having $R_e < 20$ pixels. 
The overall trends are very similar to those in Fig. 7.
\label{fig12}}
\end{figure*}
\begin{figure*}
\figurenum{13{\it a}}
\centerline{\psfig{file=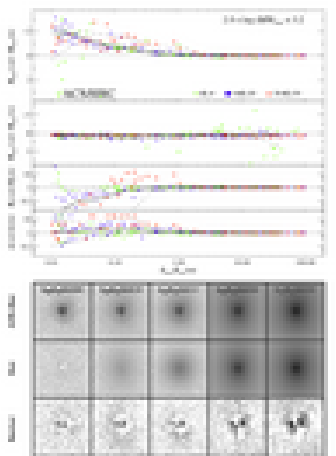,width=18cm,angle=0}}
\figcaption[fig13a.ps]
{
Simulation results for the ACS/HRC. The overall trends are 
quite similar to those for WF3 on WFPC2. See Fig. 6 for details.
\label{fig13a}}
\end{figure*}
\begin{figure*}
\figurenum{13{\it b}}
\centerline{\psfig{file=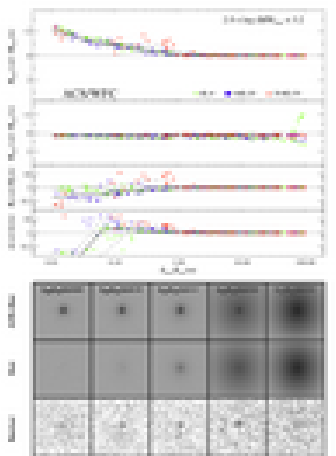,width=18cm,angle=0}}
\figcaption[fig13b.ps]
{Simulation results for ACS/WFC. The overall trends are  
quite similar to those for WF3 on WFPC2. See Fig. 6 for details.
\label{fig13b}}
\end{figure*}
\begin{figure*}
\figurenum{13{\it c}}
\centerline{\psfig{file=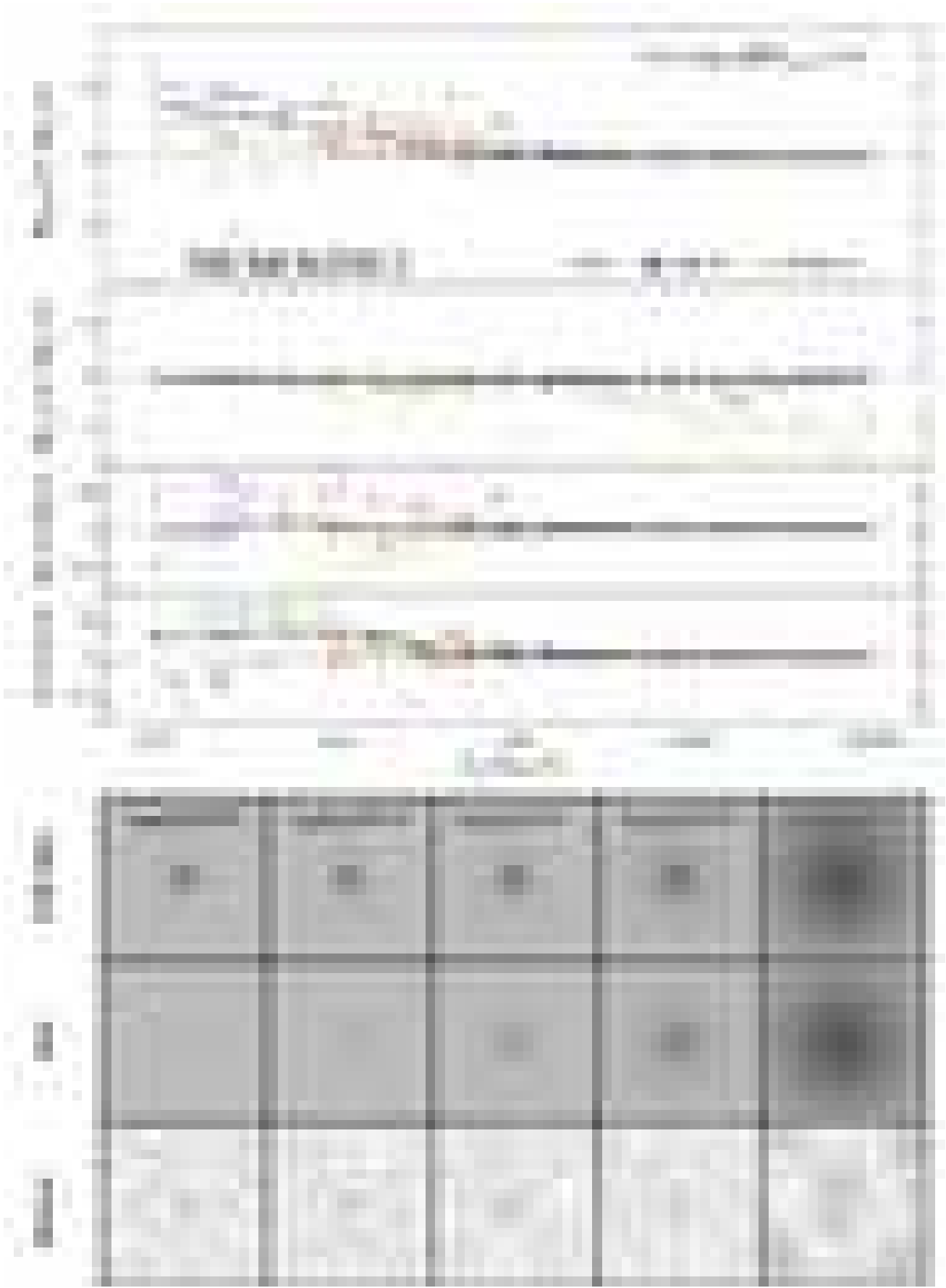,width=18cm,angle=0}}
\figcaption[fig13c.ps]
{Simulation results for NICMOS/NIC2. The overall trends are  
quite similar to those for WF3 on WFPC2. See Fig. 6 for details.
\label{fig13c}}
\end{figure*}
\begin{figure*}
\figurenum{13{\it d}}
\centerline{\psfig{file=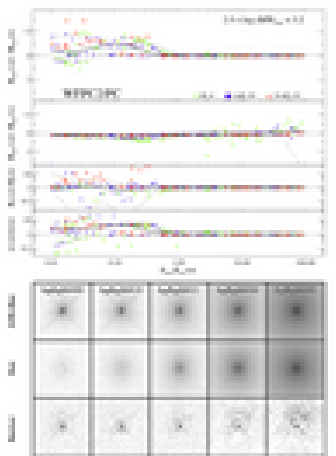,width=18cm,angle=0}}
\figcaption[fig13d.ps]
{Simulation results for the WFPC2/PC. The overall trends are  
quite similar to those for WF3 on WFPC2. See Fig. 6 for details.
\label{fig13d}}
\end{figure*}

\vspace*{+0.3cm}

\subsection{Simulating Other Detectors Onboard HST}

Different detectors onboard \hst\ have different characteristics, and so we 
further perform simulations for each of them.  As before, for each detector we 
use two different stellar PSFs to account for the PSF mismatch.  For 
simplicity, the simulations are done with 1000 samples for 
$10^{2.8}<$\nsn$<10^{3.2}$.  The detectors assumed are the Planetary Camera 
(PC) of WFPC2, the High Resolution Camera (HRC) of the Advanced Camera for 
Surveys (ACS), the Wide-Field Camera (WFC) on ACS, and NIC2 of Near-Infrared 
Camera and Multi-Object Spectrometer (NICMOS).  These are the most often-used 
detectors for AGN imaging.  Figure 13 illustrates that, although the results 
for the various detectors are slightly different, the overall trends are 
similar to those of WF3.  As shown in the previous experiments (Fig.  7), at 
\hnr$<1$ the host luminosities are systematically overestimated due to the PSF 
mismatch, and the nuclear luminosities tend to be underestimated due to the 
undersampled PSF.  Not surprisingly, the nuclear luminosities are well 
recovered in the experiments with NIC2 because it has the least undersampled 
PSF (FWHM $\approx$ 2 pixels).  We also perform the same test with reversing 
the role of PSFs as discussed in \S{3.6}. The overall trends barely 
change, although the scatter is slightly affected by this test.

\subsection{Complications}

In addition to the situations presented above, additional complications
involve extracting hosts beneath saturated AGNs or performing bulge-to-disk
decomposition on AGN hosts.  However, characterizing parameter uncertainties 
in these situations, with considerably larger parameter spaces, is beyond the
scope of this study.  Nevertheless, we conduct a brief simulation to test how
well the parameters can be recovered under very specific circumstances
corresponding to single-orbit, WF3 data, with \nsn\ $\approx 1000$.  The 
simulations below are otherwise created exactly like previous ones, 
with the reference PSFs being different than the one used to fit the model 
images, which are Nyquist-sampled. 

\bigskip
{\it Saturated AGN.}\ \ \ \ \ To test instances when the AGN is saturated, we
mask out the central few saturated pixels in the artificial images.  We create 
a set of artificial images with the \ser\ indices of the host varying from 
$n = 1$ to $n = 6$. 
With the nucleus masked out, there is 
little leverage to reliably determine the host galaxy central concentration. 
We thus hold the \ser\ index of the host galaxy fixed to $n = 4$ to fit the 
images.  In so doing, we find that the resultant luminosity extracted for the 
host galaxy has a scatter that is 0.2 mag larger compared to the unsaturated 
cases. However, with the exception of the scatter, the overall distributions 
from both tests are in good agreement. 

One can achieve high-dynamic range imaging of AGN hosts without saturating the 
core by combining a short exposure of the center with a long exposure of the 
outskirts of the galaxy.  In this circumstance, however, the noise properties
across the image will not be uniform, a situation not captured in our 
simulations.  Nevertheless, as we have shown, the systematic errors of the 
fits are almost always dominated by PSF mismatch rather than by Poisson noise.

\bigskip
{\it Bulge-to-disk-to-AGN Decomposition.}\ \ \ \ \ Characterizing measurement
uncertainties of bulge-to-disk-to-AGN (B/D/A) decomposition involves an
additional 7 free parameters, if the bulge is a \ser\ profile and is an
independent component.  This set of simulations again corresponds to \nsn\
$\approx 1000$; B/D/A decompositions can only be done on objects with high 
$S/N$.  In the interest of characterizing the uncertainties in measuring the 
bulge component in isolation, we create two sets of artificial images. The 
first corresponds to a set of
pure bulges that we define as the control sample, for which we only do a
bulge-to-AGN (B/A) decomposition.  The second set takes the same bulges,
around which we assign an exponential disk.  In so doing, uncertainties in
bulge measurements from the B/D/A decomposition can be directly compared with
the control sample of B/A decomposition.  

For the bulge components, the \ser\ indices lie in the range $1 \leq n \leq 6$.
The reference disk models are pure exponentials and have the following 
parameters:  disk scalelength $(1.5-6)\,R_{e}$ and disk luminosities
$(0.2-7)\,L_{\rm bul}$.
To fit the model images, we fix $n=4$ for the bulge and $n=1$ for the disk. In 
so doing, we find that the measurement errors for the bulge luminosity for both
samples are comparable (Fig. 14) when the bulges have intrinsic \ser\ indices
$n\gtrsim2$.  On the other hand, if the spheroid is inherently a pseudobulge 
(Kormendy \& Kennicutt 2004), where $n$ is smaller than 2, the extracted bulge 
luminosity is significantly overestimated by roughly 0.4 mag. Also, this test 
sometimes leaves substantial outliers ($\geq 0.5$ mag). The overall scatter in 
the bulge luminosity is 0.18 mag for the control sample of pure bulge 
systems, whereas the scatter grows to 0.28 mag for the later-type galaxies 
that have both a bulge and a disk. Thus, we conclude that the 
measurement error for the bulge luminosity increases by 0.1 mag in cases 
\psfig{file=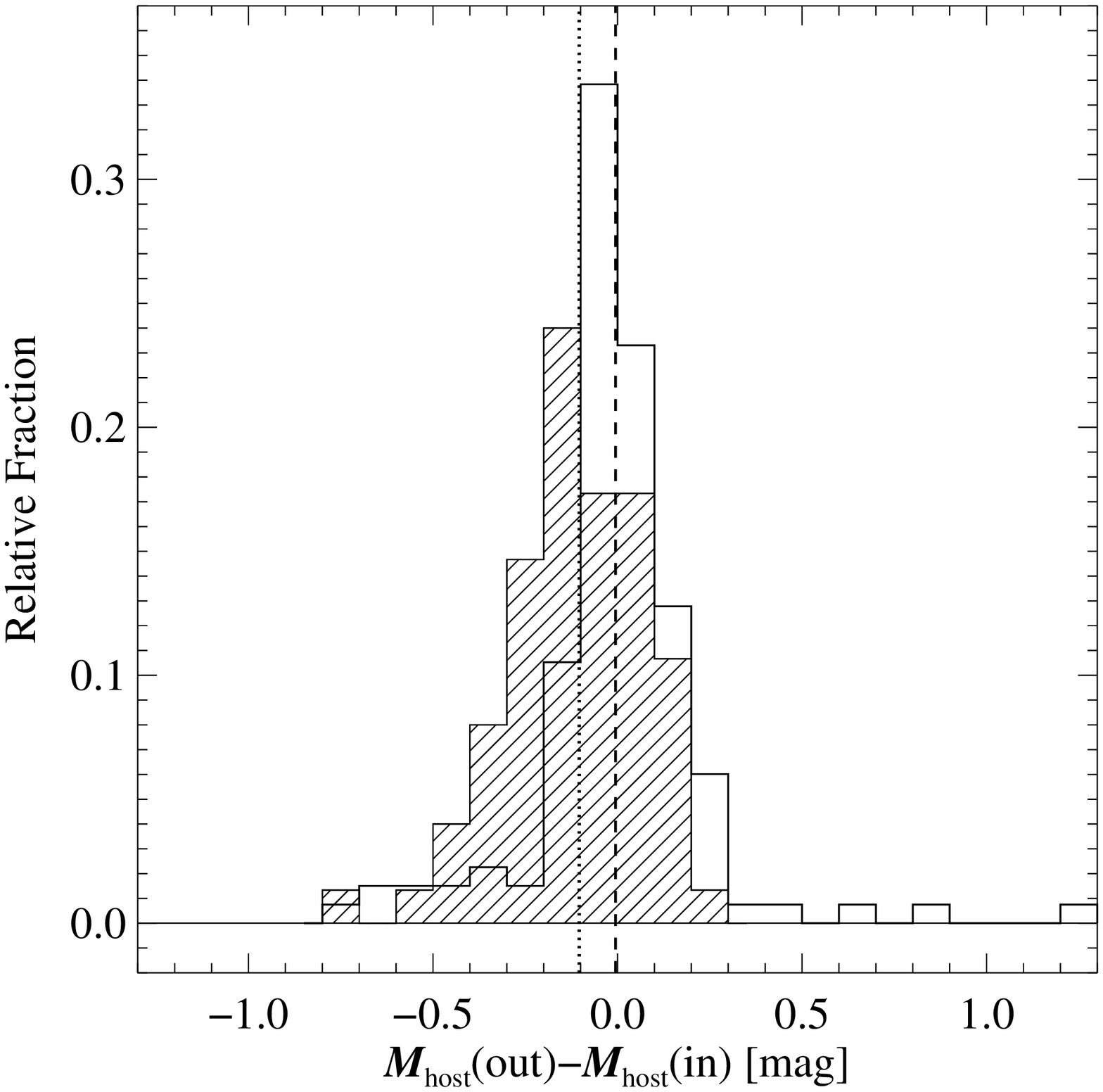,width=8.5cm,angle=0}
\figurenum{14}
\figcaption[fig14.ps]
{
Distribution of residuals for the bulge magnitudes. Here we show the fitting
results for pure bulge systems ({\it hatched histogram}) with the median value
marked with a {\it dotted line}.  The later-type galaxies that contain both a
bulge and a disk are plotted as an {\it open histogram}, and their
corresponding median value is marked with a {\it dashed line}.
\label{fig14}}
\vskip 0.3cm
\noindent
where B/D/A decomposition is required.  

\section{Discussion}

In this study, we performed 2-D image-fitting simulations of AGN
hosts to illustrate how systematics in the PSF mismatch may affect the
deblending of the AGN and the host galaxy components.  Based on these 
simulations, we suggest practical strategies for how to observe and analyze 
host galaxies with bright active nuclei.

As seen in Figures 6--7, careful determination of the PSF is needed to extract
very high-contrast subcomponents accurately.  This point was also underscored 
in the study of Hutchings et al. (2002).  We find that the three factors
that most affect PSF structures are spatial distortion, temporal changes,
and pixel 
undersampling.  The SED of the PSF matters to a much lesser degree,
but it is worthwhile to match it when possible.  The spatial variation can be
reduced by observing a stellar PSF at the same position as the science images.
However, the temporal variance is trickier to avoid without observing PSFs
concurrently with the science data, which is observationally expensive and 
rarely done in practice.  Even
then, some temporal variability may happen even within a single orbit.
Nevertheless, as there is evidence that PSF mismatch grows over time, and
may not be completely periodic in nature, it is important to obtain a stellar
PSF as close in time as possible to the science images.  
From our tests (\S{2.1}), we recommend that stellar PSF images be taken 
within a month of the science images.  In the absence of a well-matched 
stellar PSF, synthetic PSFs generated with TinyTim are adequate substitutes 
(compare Fig. 10 to Fig. 7 and Fig. 9).

In all scenarios, whether using stellar or TinyTim PSFs, both the science
image and the PSF image should be oversampled to reduce errors caused by pixel
undersampling and subpixel shifting.  This can be accomplished by observing the 
science image and the PSF star using a four-point ``dither'' pattern to recover
finer pixel sampling.  Or, if this option is not available, then 
simply broadening the images through convolution during the analysis stage is 
an acceptable alternative solution, and certainly better than no treatment at 
all (\S{3.4} and Fig. 10).

%
\begin{figure*}
\figurenum{15}
\psfig{file=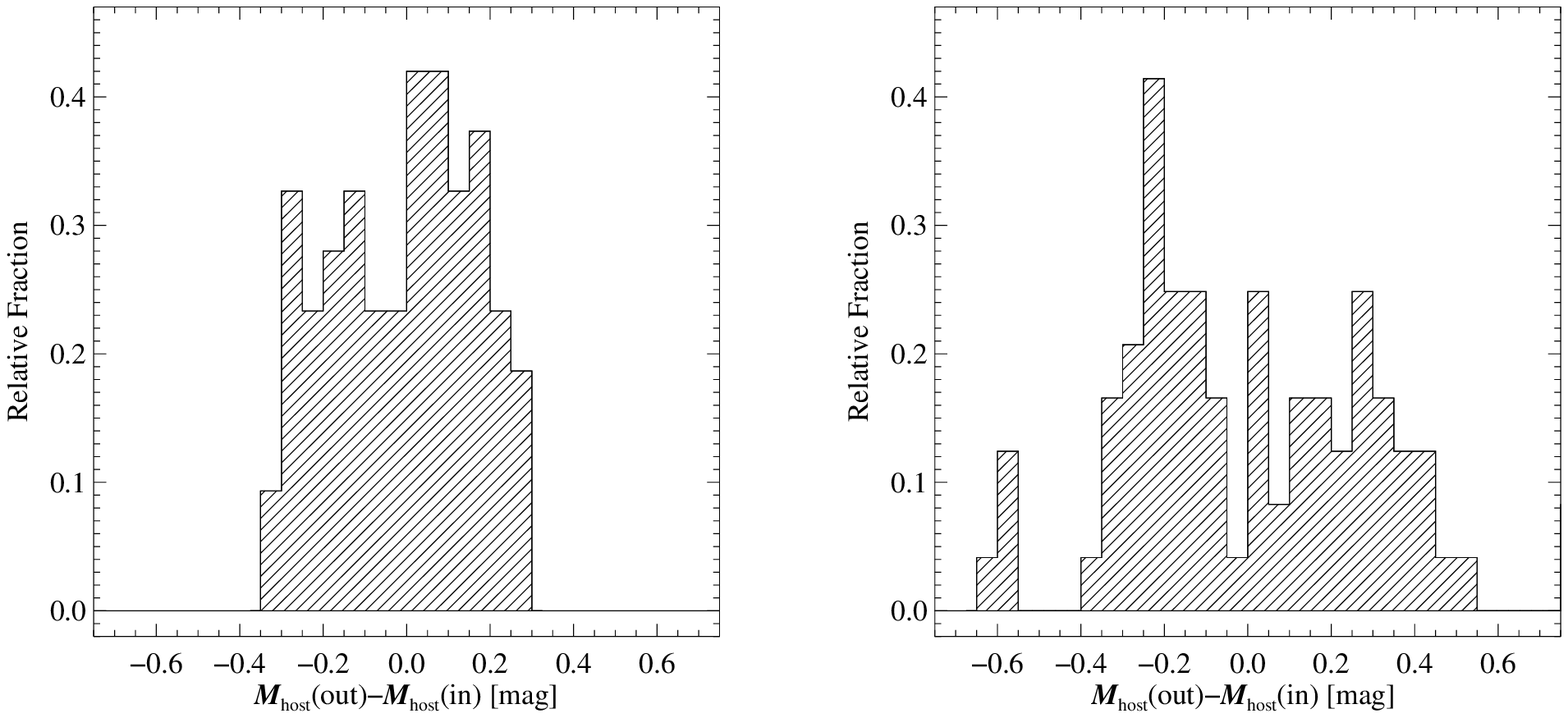,width=18.5cm,angle=0}
\figcaption[fig15.ps]
{Distribution of residuals for the magnitudes of the host for ({\it left}) 
\hnr $\geq 0.2$ and ({\it right}) \hnr $< 0.2$.  After producing artificial 
images by varying $n$ between 2 and 6, we fit them with  $n$ fixed to 4.  This 
simulation gives an estimate of the measurement error for the bulge luminosity. 
\label{fig15}}
\end{figure*}

When the host galaxies are faint and difficult to fit, it may be useful to
hold the \ser\ index fixed to a constant value of either $n=1$ or $n=4$.  The
decision about what value to use might be based on visual morphology, or by
comparing $\chi^2$ values of the bimodal priors.  A similar conclusion was
reached by McLure et al. (1999), Jahnke et al.  (2004), and S\'{a}nchez et 
al.  (2004) in their analysis of quasar host galaxies.  By varying the \ser\ 
index between $n$ = 2 and $n$ = 6, we find that the scatter is $\sigma \approx 
0.3 $ mag for \hnr $\geq 0.2$ and \nsn\ $\approx 1000$ (Fig. 15).  At higher 
contrast, \hnr $\leq 0.2$, the scatter increases to $\sigma \approx 0.4 $ mag.

Lastly, we briefly conducted a three-component bulge/disk/AGN decomposition to
characterize uncertainties in estimating the bulge component.  We find that
when the $S/N$ is high, the contrast is sufficiently low, and the bulges are
sufficiently well-resolved, then a B/D/A decomposition can yield reliable
bulge luminosity measurements.  These simulations are fully compatible with
images of $z\lesssim 0.3$ quasar host galaxies (e.g., McLure \& Dunlop 1999). 
Our tests suggest that, when three-component decomposition is required, 
the uncertainty in the bulge luminosity increases
by an additional $\sim$0.1 mag compared to fits without a disk component.

Finally, we note in passing that our simulations allow the sky value 
to vary during the fitting.  For the current application, this choice makes
little difference because the images have a large sky area and the 
profiles are idealized.  However, in real science images, the 
situation will be different.  From prior experience with actual \hst\ data, 
keeping the sky value fixed prevents the \ser\ index from going up
to extremely high values in situations where the light profile of the
galaxy is not well represented by the \ser\ function.  Thus, we recommend
keeping the sky parameter fixed to a well-determined value whenever possible.

\acknowledgements

This work was supported by the Carnegie Institution of Washington and by NASA 
grants HST-AR-10969, HST-GO-10149, and HST-GO-10395 from the Space Telescope 
Science Institute (operated by AURA, Inc., under NASA contract NAS5-26555).
M.K. and M.I. acknowldege the support of the Korea Science  and Engineering 
Foundation (KOSEF) through the Astrophysical Research  Center for the 
Structure and Evolution of the Cosmos (ARCSEC).
C.Y.P. was supported through the Institute/Giacconi Fellowship (STScI) and the
Plaskett Fellowship (NRC-HIA, Canada). Research by A.J.B. is supported by NSF 
grant AST-0548198. 
We acknowledge the useful suggestions of the referee.
We thank Paul Martini for helpful comments on the 
manuscript.

\appendix
\section{Simulation Results as a Function of \nsn}
Here we examine how the fitting results depend on the $S/N$ of the nucleus. 
Figures 16--18 show the simulation results as a function of \nsn. We divide 
the sample into 6 bins in \nsn\ and 3 bins according to the PSF used for the 
fit.

\clearpage

\begin{figure*}
\figurenum{16{\it a}}
\psfig{file=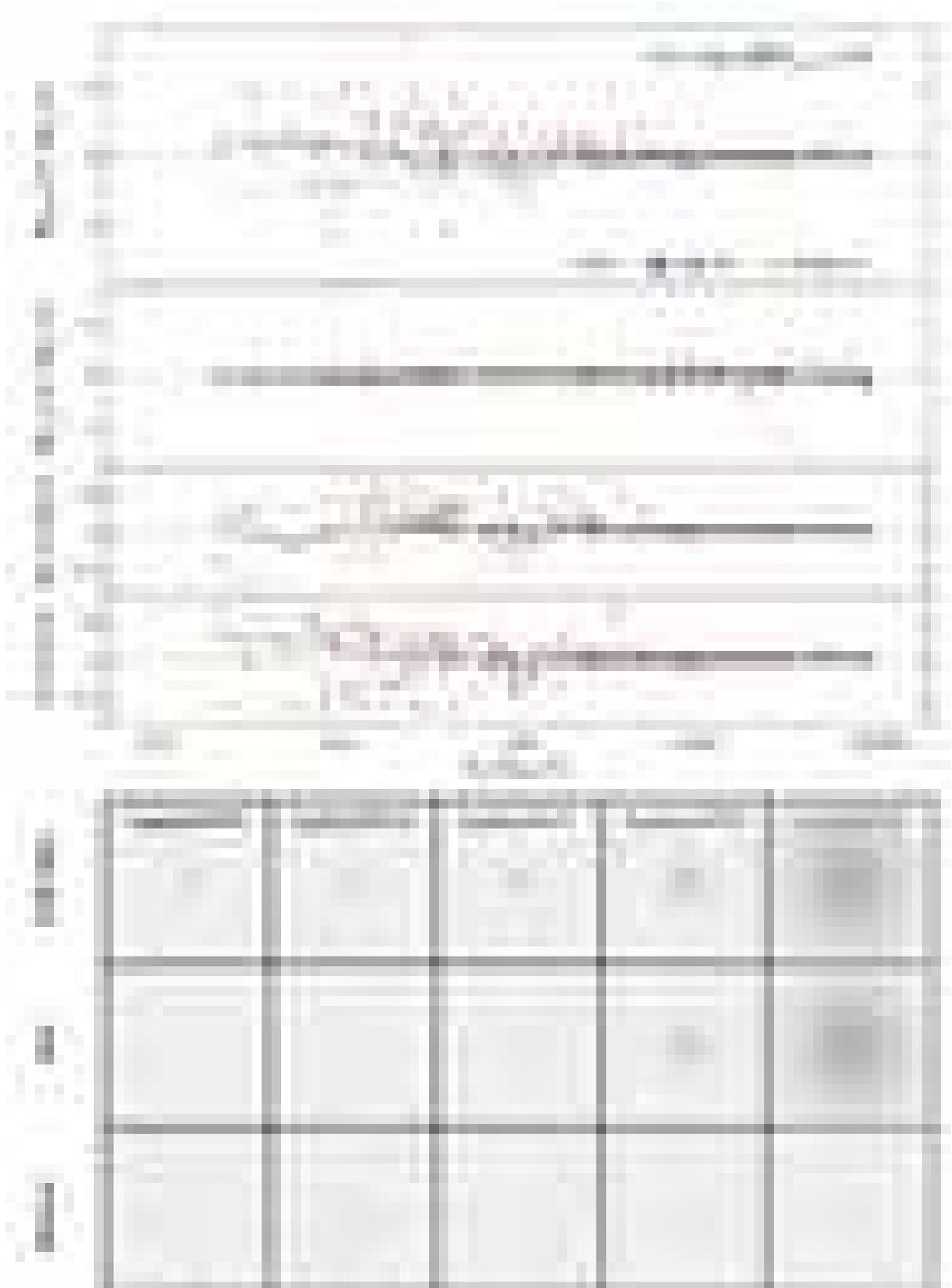,width=18.5cm,angle=0}
\figcaption[fig16a]
{
Simulation results for the idealized situation in which the fitting PSF is
identical to that used for generating the input images; see Fig. 6 for
details. We display $10^{1.6} \leq$ \nsn\ $\leq 10^{2.0}$.  
\label{fig16a}}
\end{figure*}

\begin{figure*}
\figurenum{16{\it b}}
\psfig{file=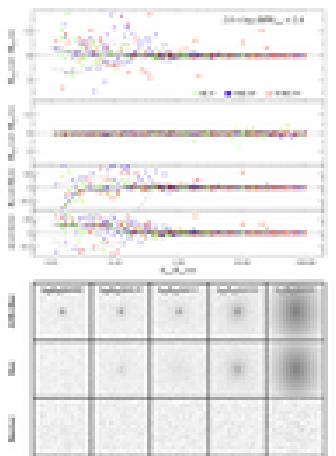,width=18.5cm,angle=0}
\figcaption{
Simulation results for $10^{2.0} \leq$ \nsn\ $\leq 10^{2.4}$.
}
\end{figure*}

\begin{figure*}
\figurenum{16{\it c}}
\psfig{file=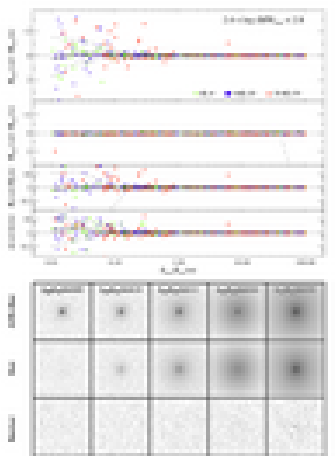,width=18.5cm,angle=0}
\figcaption{
Simulation results for $10^{2.4} \leq$ \nsn\ $\leq 10^{2.8}$.
}
\end{figure*}

\begin{figure*}
\figurenum{16{\it d}}
\psfig{file=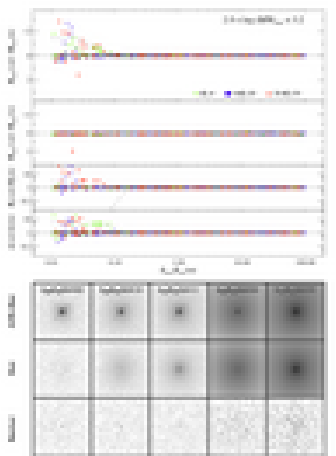,width=18.5cm,angle=0}
\figcaption{
Simulation results for $10^{2.8} \leq$ \nsn\ $\leq 10^{3.2}$.
}
\end{figure*}

\begin{figure*}
\figurenum{16{\it e}}
\psfig{file=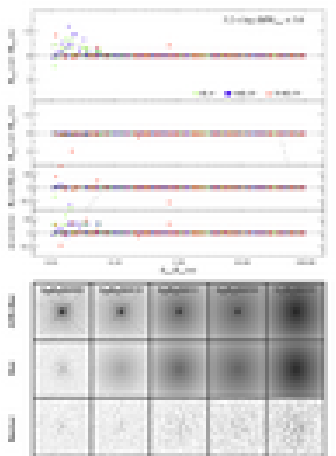,width=18.5cm,angle=0}
\figcaption{
Simulation results for $10^{3.2} \leq$ \nsn\ $\leq 10^{3.6}$.
}
\end{figure*}

\begin{figure*}
\figurenum{16{\it f}}
\psfig{file=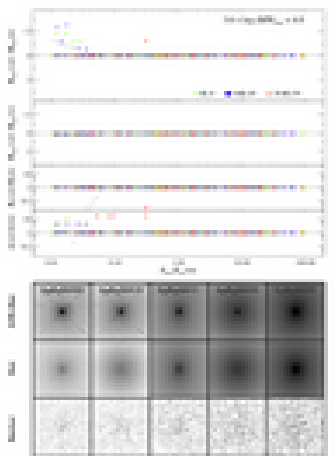,width=18.5cm,angle=0}
\figcaption{
Simulation results for $10^{3.6} \leq$ \nsn\ $\leq 10^{4.0}$.
}
\end{figure*}

\begin{figure*}
\figurenum{17{\it a}}
\psfig{file=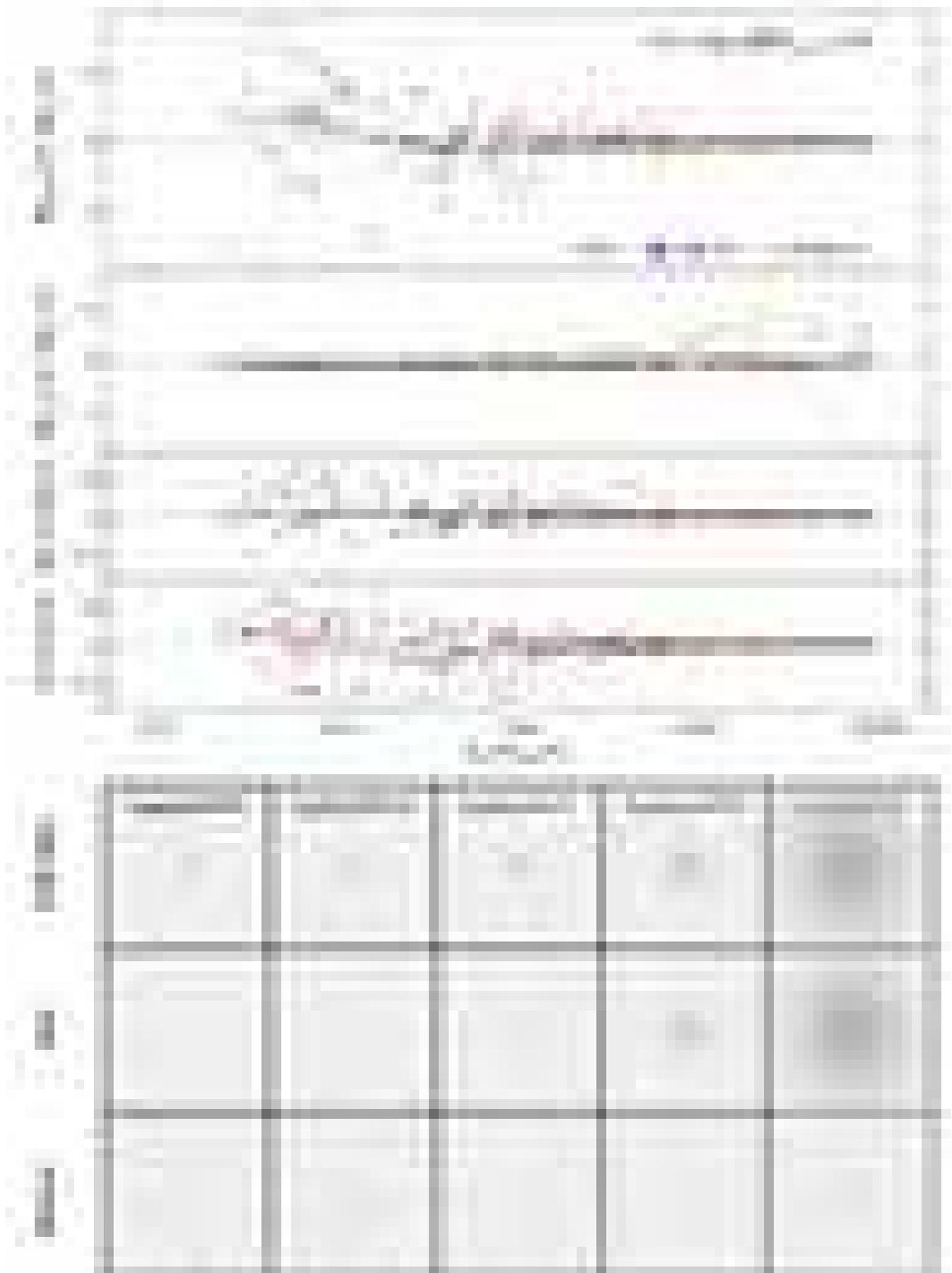,width=18.5cm,angle=0}
\figcaption[fig16a]
{
Similar to Fig. 16{\it a}, except that here we fit the artificial images
with a PSF different from the one used for generating the input images;
see Fig. 6 for details. We display $10^{1.6} \leq$ \nsn\ $\leq 10^{2.0}$.
\label{fig17a}}
\end{figure*}

\begin{figure*}
\figurenum{17{\it b}}
\psfig{file=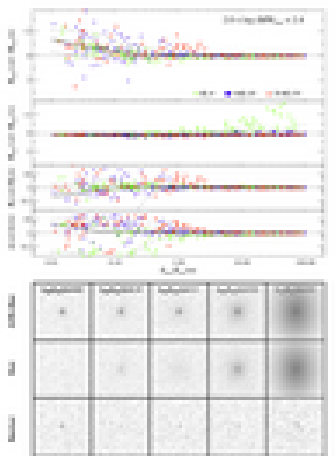,width=18.5cm,angle=0}
\figcaption{
Simulation results for $10^{2.0} \leq$ \nsn\ $\leq 10^{2.4}$.
}
\end{figure*}

\begin{figure*}
\figurenum{17{\it c}}
\psfig{file=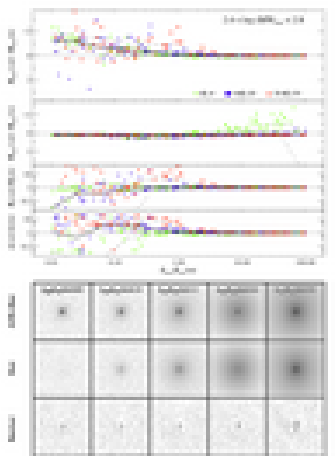,width=18.5cm,angle=0}
\figcaption{
Simulation results for $10^{2.4} \leq$ \nsn\ $\leq 10^{2.8}$.
}
\end{figure*}

\begin{figure*}
\figurenum{17{\it d}}
\psfig{file=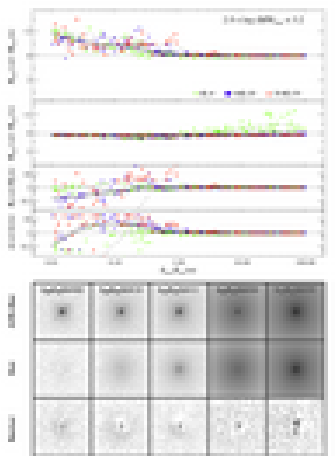,width=18.5cm,angle=0}
\figcaption{
Simulation results for $10^{2.8} \leq$ \nsn\ $\leq 10^{3.2}$.
}
\end{figure*}

\begin{figure*}
\figurenum{17{\it e}}
\psfig{file=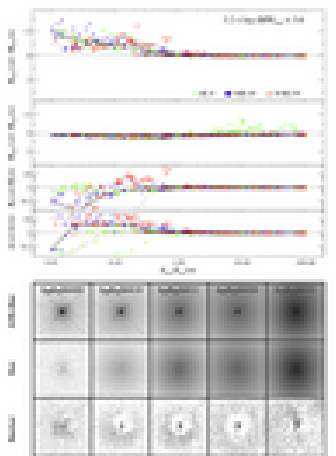,width=18.5cm,angle=0}
\figcaption{
Simulation results for $10^{3.2} \leq$ \nsn\ $\leq 10^{3.6}$.
}
\end{figure*}

\begin{figure*}
\figurenum{17{\it f}}
\psfig{file=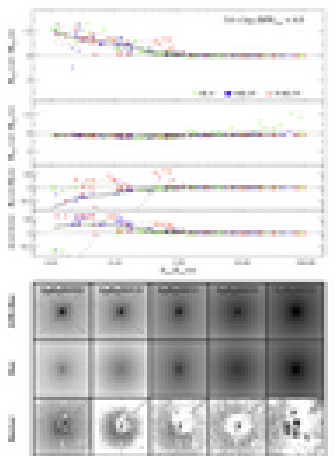,width=18.5cm,angle=0}
\figcaption{
Simulation results for $10^{3.6} \leq$ \nsn\ $\leq 10^{4.0}$.
}
\end{figure*}

\begin{figure*}
\figurenum{18{\it a}}
\psfig{file=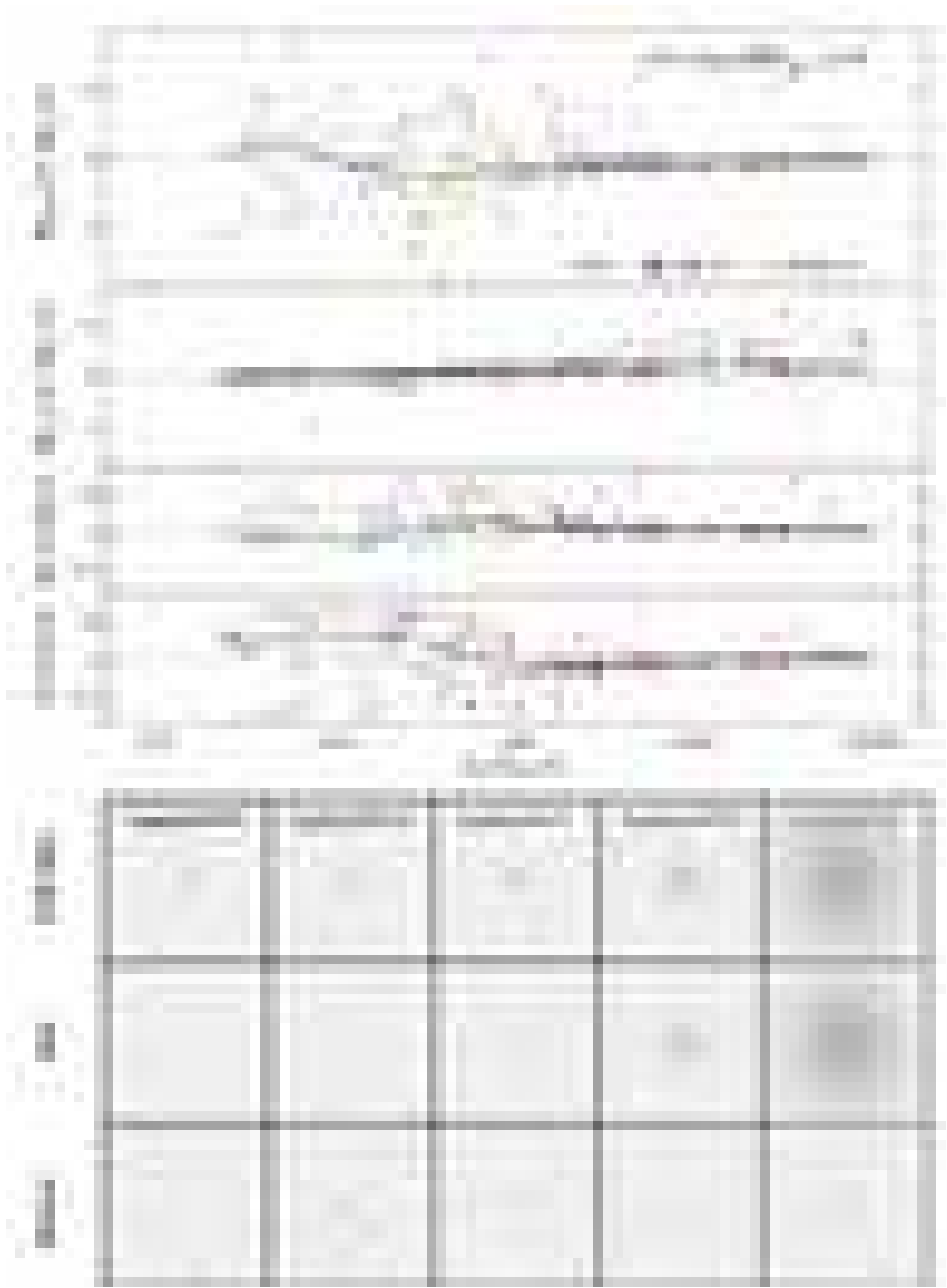,width=18.5cm,angle=0}
\figcaption[fig16a]
{
Similar to Fig. 16{\it a}, except that here we fit the images with the TinyTim
PSF; see Fig. 6 for details. We display $10^{1.6} \leq$ \nsn\ $\leq 10^{2.0}$.
\label{fig18a}}
\end{figure*}

\begin{figure*}
\figurenum{18{\it b}}
\psfig{file=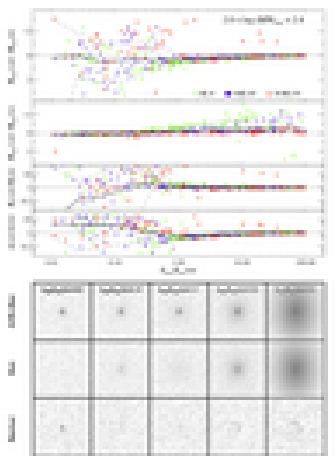,width=18.5cm,angle=0}
\figcaption{
Simulation results for $10^{2.0} \leq$ \nsn\ $\leq 10^{2.4}$.
}
\end{figure*}

\begin{figure*}
\figurenum{18{\it c}}
\psfig{file=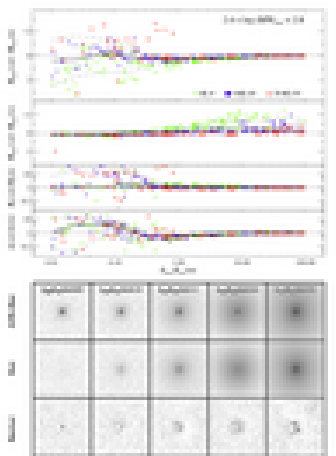,width=18.5cm,angle=0}
\figcaption{
Simulation results for $10^{2.4} \leq$ \nsn\ $\leq 10^{2.8}$.
}
\end{figure*}

\begin{figure*}
\figurenum{18{\it d}}
\psfig{file=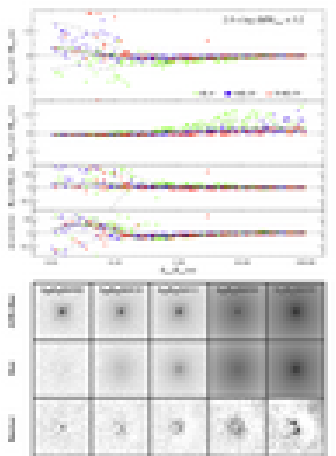,width=18.5cm,angle=0}
\figcaption{
Simulation results for $10^{2.8} \leq$ \nsn\ $\leq 10^{3.2}$.
}
\end{figure*}

\begin{figure*}
\figurenum{18{\it e}}
\psfig{file=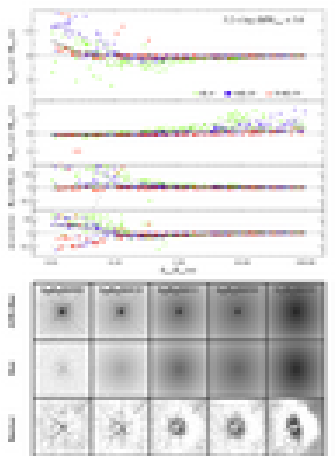,width=18.5cm,angle=0}
\figcaption{
Simulation results for $10^{3.2} \leq$ \nsn\ $\leq 10^{3.6}$.
}
\end{figure*}

\begin{figure*}
\figurenum{18{\it f}}
\psfig{file=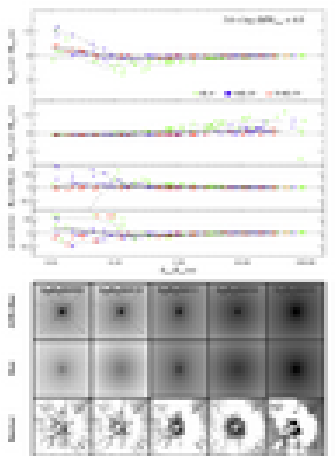,width=18.5cm,angle=0}
\figcaption{
Simulation results for $10^{3.6} \leq$ \nsn\ $\leq 10^{4.0}$.
}
\end{figure*}

\end{document}